\documentclass[10pt,twocolumn,twoside]{IEEEtran}

\pdfoutput=1
\usepackage{amsmath}
\usepackage{amssymb}
\usepackage{graphicx}
\usepackage{tikz}
\usepackage{algorithm}
\usepackage{algorithmic}
\usepackage{hhline}
\usepackage{booktabs}
\usepackage{array}
\usepackage{multirow}
\usepackage{url}
\usepackage{setspace}

\usetikzlibrary{fit,positioning}

\definecolor{grey30}{RGB}{84,84,84}
\definecolor{ored}{RGB}{255,36,0}
\definecolor{metal}{RGB}{35,107,142}
\definecolor{dsblue}{RGB}{0,104,139}

\newcommand{\symvec}[1]{{\mbox{\boldmath $#1$}}}
\newcommand{\symmat}[1]{{\mbox{\boldmath $#1$}}}
\newcommand{\var}{\textrm{v}}
\newcommand{\hermitian}{{\textrm{H}}}
\usepackage{cleveref}
\crefformat{footnote}{#2\footnotemark[#1]#3}

\usepackage[final]{review}

\begin{document}
\title{A Variational EM Algorithm for the Separation of Time-Varying Convolutive Audio Mixtures}

\author{Dionyssos Kounades-Bastian,
        Laurent~Girin,
        Xavier~Alameda-Pineda,
        Sharon Gannot,
        Radu Horaud
\thanks{D. Kounades-Bastian is with INRIA Grenoble Rh\^one-Alpes, France. E-mail: \texttt{dionyssos.kounades-bastian@inria.fr}}
\thanks{L. Girin is with INRIA Grenoble Rh\^one-Alpes, France, and with Univ. Grenoble Alpes, GIPSA-lab, Grenoble, France. E-mail: \texttt{laurent.girin@gipsa-lab.grenoble-inp.fr}}%
\thanks{X. Alameda-Pineda is with University of Trento, Italy. E-mail: \texttt{xavier.alamedapineda@unitn.it}}
\thanks{S. Gannot is with Bar Ilan University, Faculty of Engineering, Israel. E-mail: \texttt{Sharon.Gannot@biu.ac.il}}
\thanks{R. Horaud is with INRIA Grenoble Rh\^one-Alpes, France. E-mail: \texttt{radu.horaud@inria.fr}}
\thanks{D. Kounades-Bastian, L. Girin and R. Horaud acknowledge support from the European FP7 STREP project EARS \#609465 and from 
the European Research Council through the ERC Advanced Grant VHIA \#340113.}
}

\maketitle

\begin{abstract}
This paper addresses the problem of separating audio sources from time-varying convolutive mixtures. We propose a probabilistic 
framework based on the local complex-Gaussian model combined with non-negative matrix factorization. The time-varying mixing filters are modeled by a continuous temporal stochastic process. 
We present a variational expectation-maximization (VEM) algorithm that employs a Kalman smoother to estimate the time-varying mixing matrix, and that jointly estimate the source parameters. 
The sound sources are then separated by Wiener filters constructed with the estimators provided by the VEM algorithm.
Extensive experiments on simulated data show that the proposed method outperforms a block-wise version of a state-of-the-art baseline method. 
\end{abstract}


\ifCLASSOPTIONcaptionsoff
  \newpage
\fi

\section{Introduction}
\label{sec:intro}

Source separation aims at recovering unobserved source signals from observed mixtures \cite{comon10ha}.
Audio source separation (ASS) is mainly concerned with mixtures of speech, music, ambient noise, etc. 
For acoustic signals in natural environments, the mixing process is generally considered as \emph{convolutive}, 
i.e., the acoustic channel between each source and each microphone is modeled by a linear filter that represents the multiple source-to-microphone paths due to reverberations. 
Source separation is a major component of machine audition systems, since it is used as a preprocessing step for many higher-level processes such as speech recognition, human-computer or human-robot interaction. 


The vast majority of works on ASS from convolutive mixtures deals with \emph{time-invariant} mixing filters, which means that the position of sources and microphones is assumed to be fixed. In other words, the source-to-microphone acoustic paths are assumed to remain the same over the duration of the recordings. In this work we consider the more realistic case of \emph{time-varying} convolutive mixtures corresponding to source-to-microphone channels that can change over time. This should be able to take into account possible source or microphone motions. For example, in many Human-robot interaction scenarios, there is a strong need to consider mixed speech signals emitted by \textit{moving} speakers, and/or recorded by a \textit{moving} robot, and perturbed by reverberations. More generally, changes in the environment such as door/window opening/closing or curtain pulling must also be accounted for. 
Note that in this paper, the mixtures under consideration can be \emph{underdetermined}, i.e., there may be less microphones than sources, which is a difficult ASS problem in its own right \cite{comon10ha}.

\subsection{Related Work}
The ASS literature that deals with time-invariant mixing filters is much larger than the literature dealing with time-varying filters. Therefore, we briefly discuss the former before reviewing the latter. State-of-the-art time-invariant ASS methods generally start with a time-frequency (TF) decomposition of the temporal signals, e.g., by applying the short-time Fourier transform (STFT). In the TF domain, the time-invariant convolutive filters are converted to multiplicative coefficients independent at each frequency bin \cite{avargel07mu}. These methods can then be classified into three (non-exclusive) categories \cite{vincent10pr}. Firstly, separation methods based on independent component 
analysis (ICA) consist in estimating the demixing filters that maximize the independency of separated sources \cite{comon10ha,hyvarinen01in}. Unfortunately, ICA-based 
methods are subject to the well-known scale ambiguity and source permutation problems across frequency bins. In addition, these methods cannot be applied to underdetermined mixtures. Secondly, methods based on sparse component analysis (SCA) and binary masking rely on the assumption that only one source is active at each TF point \cite{Win07, mandel10mo}.
Thirdly, more recent methods are based on complex-valued local Gaussian models (LGMs) for the sources 
\cite{liutkus11ga}, and the model proposed here is a member of this family of methods. 

The LGM was initially proposed for single-microphone speech enhancement \cite{ephraim84sp}, 
then extended to single-channel ASS \cite{benaroya03no,benaroya06au} and multi-channel ASS 
\cite{fevotte05ma,ozerov10mu,duong10un,ozerov12ag}. The method proposed in \cite{ozerov10mu} 
provides a rigorous framework for ASS from underdetermined convolutive mixtures: An LGM source model is combined with a nonnegative matrix factorization (NMF)
model \cite{lee99le, lee01al} applied to the source PSD matrix \cite{fevotte09no}, which is reminiscent of pioneering works such as \cite{benaroya03no}. This allows one to drastically reduce the number of model parameters and to alleviate the source permutation problem. However, in \cite{ozerov10mu} the mixing filters do not vary over time:  they are considered as model parameters and, together with the NMF coefficients, they are estimated via an EM algorithm. Then, the sound sources are separated with Wiener filters constructed from the learned parameters. A similar LGM-based approach is adopted in \cite{yoshioka11bl}, though the speech signal PSD is here modeled as a time-varying auto-regressive (AR) model. Here also, all model parameters are estimated by maximizing the likelihood of the observed signals and solved by EM iterations.

In comparison to the time-invariant methods that we just mentioned, the literature dealing with time-varying acoustic mixtures is scarce.
Early attempts addressing the separation of time-varying mixtures basically consisted in block-wise adaptations of time-invariant methods: 
An STFT frame sequence is split into blocks, and a time-invariant ASS algorithm is applied to each block.
Hence, block-wise adaptations assume time-invariant filters within blocks. The separation parameters are updated from one block to the next and the separation result over a 
block can be used to initialize the separation of the next block. 
Frame-wise algorithms can be considered as particular cases of block-wise algorithms, with single-frame blocks, 
and hybrid methods may combine block-wise and frame-wise processing. Notice that, depending on the implementation, some of these methods may run online.

Interestingly, most of the block-wise approaches use ICA, either in the temporal domain
\cite{anemuller99on} (limited to anechoic setups), \cite{koutras00bl, hild02bl, aichner03on, prieto05bl} or in the Fourier domain \cite{mukai03ro}, \cite{addison06bl} (limited to instantaneous mixtures), \cite{nakadai09so}. In addition to being limited to overdetermined mixtures, block-wise ICA methods need to account for the source permutation problem, not only across frequency bins, as usual, but across successive blocks as well. Examples of block-wise adaptation of binary-masking or LGM-based methods are more scarce. As for binary 
masking, a block-wise adaptation of \cite{araki07un} is proposed in \cite{loesch09on}. This method performs source separation by 
clustering the observation vectors in the source image space.
As for LGM, \cite{simon12ag} describes an online block- and frame-wise adaptation of the general LGM framework proposed in \cite{ozerov12ag}. 
One important problem, common to all block-wise approaches, is the difficulty to choose the block size. Indeed, the block 
size must assume a good trade-off between local channel stationarity (short blocks) and sufficient data to infer relevant statistics (long blocks). 
The latter constraint can drastically limit the dynamics of either the sources or the sensors \cite{loesch09on}. Other parameters such as the step-size of the 
iterative update equations may also be difficult to set \cite{simon12ag}. In general, 
systematic convergence towards a good separation solution using a limited amount of signal statistics remains an open issue.

Dynamic scenarios were also addressed differently in \cite{markovich10su}, where a beamforming method for extracting multiple moving sources is proposed. This method is applicable only to over-determined mixture. Also, iterative and sequential approaches for speech enhancement in reverberant environment were proposed in \cite{weinstein94it}. The proposed methods utilize the EM framework to jointly estimate the desired speech signal and the required (deterministic) parameters, namely the speech AR coefficients, and the speech and noise mixing filters taps. For on-line implementation, a recursive version of the M-step was developed and the Kalman smoother, used in the batch mode, is substituted by the Kalman filter. However, only the case of a $2 \times 2$ mixture was addressed.


Separating underdetermined time-varying convolutive mixtures using binary masking within a probabilistic LGM framework was proposed in 
\cite{higushi12un}. The mixing filters are considered as latent variables that follow a Gaussian distribution with mean vector depending on the 
direction of arrival (DOA) of the corresponding source. The DOA is modeled as a discrete latent variable taking values from a finite set of angles 
and following a discrete hidden Markov model (HMM). A variational expectation-maximization (VEM) algorithm is derived to perform the inference, including 
forward-backward equations to estimate the DOA sequence. This approach provides interesting results but it suffers from several limitations. First, 
the separation quality is poor, proper to binary masking approaches. Second, the accuracy is limited, which is inherent to the use of a discrete temporal model to 
represent a continuous variable, namely the source DOAs. Moreover, constraining the mixing filter to a DOA-dependent model can be problematic in 
highly reverberant environments.
Finally, it must be noted that no specific source variance model is exploited, and that the filter and DOA  models are assumed to solve the source permutation problem (both in frequency and 
time).

\subsection{Contributions}

In this paper we adopt the source LGM framework with an NMF PSD model. We consider the very general case of an underlying convolutive mixing process 
that is allowed to vary over time, and we model this process as a set of, \textit{temporally-linked continuous latent variables}, using a prior model.
We propose to parameterize the transfer function of the mixing filters with an unconstrained continuous linear dynamical system (LDS) \cite{bishop06pa}. 
We believe that this model can be more effective than the DOA-dependent HMM model of \cite{higushi12un} in adverse and reverberant conditions, since 
the relationship between the transfer function and the source DOA can be quite complex. In addition, \cite{higushi12un} relies on binary masking for 
separating the sources, which is known to introduce speech distortion, whereas we use the more general and more efficient Wiener filtering tied to 
LGM-based methods.


The proposed method may be viewed as a generalization of \cite{ozerov10mu} to moving sources, moving microphones, or both. However, exact inference 
of the posterior distribution, as proposed in \cite{ozerov10mu}, turns out to be intractable in the more general model that we consider here. 
Therefore, we propose an approximate solution for the joint estimation of the model parameters and inference of the latent variables. We derive a 
variational EM (VEM) algorithm in which a \emph{Kalman smoother} is used for the inference of the time-varying mixing filters. In comparison to the 
methodology described in \cite{simon12ag}, the proposed model goes beyond block- or frame-wise adaptation because it exploits the information 
available with the whole sequence of input mixture frames. To summarize, the proposed method exploits all the available data to estimate the source 
parameters and mixing process parameters at each frame. As a consequence, it cannot be applied online. Note that an earlier reference to the 
incorporation of a latent Bayesian continuous model into the underlying filtering, with application to speech processing, can be found 
in~\cite{gannot03on}. Two schemes were proposed, namely a dual scheme with two Kalman filters applied sequentially in parallel, and a joint scheme 
using the approximated unscented Kalman filter. Only very simple filtering schemes were addressed. In the present paper, we provide a more rigorous 
treatment of the joint signal and parameter estimation problem, using the variational approach.


This paper is an extended version of \cite{kounades15av}. A detailed description of the proposed model and of the associated VEM algorithm is now 
provided. Several mathematical derivations, that were omitted in \cite{kounades15av}, are now included in order to make the paper self-consistent, 
easy to understand, and to allow method reproducibility.  Moreover, several computational simplifications are proposed, leading to a more efficient 
implementation. The method is tested over a larger set of signals and configurations, \addnote[new-exp-intro]{1}{including experiments with blind 
initialization and real recordings}, thus extending the very preliminary results presented in \cite{kounades15av}. These results are compared with a 
block-wise implementation of the baseline method \cite{ozerov10mu}. This may well be viewed as an adaptation of the general framework \cite{simon12ag} 
to convolutive mixtures.
Matlab code of the proposed algorithm together with speech test data are provided as supplementary material.\footnote{\url{http://ieeexplore.ieee.org}}$^,$\footnote{\url{https://team.inria.fr/perception/research/vemove/}}

The remaining of the paper is organized as follows. Section \ref{sec:Models} describes the source, mixture and channel models. 
The associated VEM algorithm is described in Section~\ref{sec:varEM}. Implementation details are discussed in Section \ref{sec:Implementation}.
The experimental validation is reported in Section~\ref{sec:exp}. Conclusions and future works are discussed in Section~\ref{sec:conclusion}.


\section{Audio Mixtures with Time-Varying Filters}
\label{sec:Models}

\subsection{The Source Model}

We work in a time-frequency representation, after applying the short-time Fourier transform (STFT) to the time-domain mixture signal. Let $f \in [1,F]$ 
denote the frequency bin index, and $\ell \in [1,L]$ denote the frame index. \addnote[transpose]{1}{Consider a mixture of $J$ source signals, with
$\mathbf{s}_{f\ell}=[s_{1,f\ell} \ldots s_{J,f\ell}]^\top \in \mathbb{C}^J$ denoting the latent vector of source coefficients at 
TF bin $(f,\ell)$ ($\mathbf{x}^\top$ and $\mathbf{x}^\hermitian$ respectively denote $\mathbf{x}$ transpose and 
conjugate-transpose).}
Let $\{\mathcal{K}_{j}\}_{j=1}^J$ denote a non-trivial partition of $\{1 \ldots K\}$, $K \geq J$ (in practice we may have $K \gg J$), that is known in advance.
Following \cite{ozerov10mu}, a coefficient $s_{j,f\ell}$ is modeled as the sum of latent components $c_{k,f\ell}$,
$k \in \mathcal{K}_{j}$:

\begin{equation}
s_{j,f\ell}=\underset{k\in \mathcal{K}_{j}}\sum{c_{k,f\ell}} \Leftrightarrow \mathbf{s}_{f\ell} = \mathbf{G} \mathbf{c}_{f\ell}, \label{eq:source1}
\end{equation} 
where $\mathbf{G} \in \mathbb{N}^{J \times K}$ is a binary selection matrix with entries $G_{jk} = 1$ if $k \in \mathcal{K}_j$ and $G_{jk} = 0$ otherwise, and $\mathbf{c}_{f\ell}=[c_{1,f\ell},$ $\ldots, c_{K,f\ell}]^\top \in \mathbb{C}^K$ is the vector of component coefficients at $(f,\ell)$.
Each component $c_{k,f\ell}$ is assumed to follow a zero-mean proper complex Gaussian distribution 
with variance $w_{fk}h_{k\ell}$, where $w_{fk}, h_{k\ell}\in \mathbb{R}^+$. The components are assumed to be mutually independent and individually independent across frequency and time. 
Thus the component vector probability density function (pdf) writes:\footnote{The proper complex Gaussian distribution is defined as 
$\mathcal{N}_c(\mathbf{x};\symvec{\mu},\symmat{\Sigma}) = {|\pi \symmat{\Sigma}|^{-1}} \exp \big( - [\mathbf{x}-\symvec{\mu}]^\hermitian \symmat{\Sigma}^{-1} [\mathbf{x}-\symvec{\mu}] \big)$,
 with $\mathbf{x},\symvec{\mu} \in \mathbb{C}^I$ and $\symmat{\Sigma} \in \mathbb{C}^{I \times I}$ being the argument, mean vector, and covariance matrix respectively \cite{neeser93pr}.}
\begin{equation}\label{eq:equa1} 
p(\mathbf{c}_{f\ell}) = \mathcal{N}_{c}\Big(\mathbf{c}_{f\ell}; \mathbf{0}, \text{diag}_K \left(w_{fk}h_{k\ell} \right) \Big),
\end{equation}
where $\mathbf{0}$ denotes the zero-vector, $\text{diag}_K(d_k)$ denotes the $K \times K$ diagonal matrix with entries $[d_1 \ldots d_k \ldots 
d_K]^\top$, and the source vector pdf writes:\begin{align} \label{eq:prior_sources}
p(\mathbf{s}_{f\ell}) = \mathcal{N}_{c}\Bigg(\mathbf{s}_{f\ell}; \mathbf{0}, \text{diag}_J \bigg( \sum \limits_{k \in \mathcal{K}_j} w_{fk}h_{k\ell} \bigg)\Bigg).
\end{align}
Eq. \eqref{eq:prior_sources} corresponds to the modeling of the $F \times L$ source PSD matrix with the NMF model, which is widely used in audio analysis, audio source separation,
 and speech enhancement \cite{benaroya03no, virtanen07mo, fevotte09no, mohammadiha13su}. NMF is empirically verified to adequately model a large range of sounds by providing 
harmonic as well as non-harmonic patterns activated over time. Note that both source and component vectors are treated as latent variables linked by (\ref{eq:source1}).

\subsection{The Mixture Model}

In many source separation methods, including \cite{ozerov10mu}, the mixture signal is modeled as a time-invariant convolutive noisy mixture of the source signals.
Let us denote the $I$-channel mixture signal in the TF domain by $\mathbf{x}_{f\ell}=[x_{1,f\ell} \ldots x_{I,f\ell}]^\top \in \mathbb{C}^I$. 
Relying on the so-called narrow-band assumption (i.e. the impulse responses of the channel are shorter than the TF analysis window), 
$\mathbf{x}_{f\ell}$ writes \cite{parra00co, gannot01si}: $ \mathbf{x}_{f\ell}=\mathbf{A}_{f}\mathbf{s}_{f\ell}+\mathbf{b}_{f\ell}$, where 
$\mathbf{b}_{f\ell} = [b_{1,f\ell}  \ldots  b_{I,f\ell}]^\top \in \mathbb{C}^{I}$ is a zero-mean complex-Gaussian residual noise, and 
$\mathbf{A}_{f} = [\mathbf{a}_{1,f} \ldots  \mathbf{a}_{J,f}] \in \mathbb{C}^{I \times J}$ is the mixing matrix (a column $\mathbf{a}_{j,f} \in 
\mathbb{C}^{I}$ is the mixing vector for source $j$). This way, the mixing matrix depends only on the frequency $f$ but not on 
the time frame $\ell$, meaning that the filters are assumed to be time-invariant. 
Since we are expressly interested in modeling time-varying filters, the mixing equation naturally becomes:
\begin{equation} 
\mathbf{x}_{f\ell}=\mathbf{A}_{f\ell}\mathbf{s}_{f\ell}+\mathbf{b}_{f\ell} \label{eq:mixing},
\end{equation} 
with $\mathbf{A}_{f\ell}$ being both frequency- and time-dependent. This equation allows us to cope with possible source/sensor movements and 
other environmental changes. Note that \eqref{eq:mixing} accounts for temporal variations of the channel across frames, though it assumes that the 
channel is not varying within an individual frame, which is a reasonable assumption for a wide variety of applications.
For simplicity $\mathbf{b}_{f\ell}$ is assumed here to be stationary and isotropic, i.e. $p(\mathbf{b}_{f\ell}) = \mathcal{N}_c(\mathbf{b}_{f\ell};\mathbf{0},\var_f \mathbf{I}_{I})$, with $\var_f \in \mathbb{R}^+$ being a parameter to be estimated,
and $\mathbf{I}_{I}$ denoting the identity matrix of size $I$. 
The conditional data distribution is thus given by 
$p(\mathbf{x}_{f\ell}|\mathbf{A}_{f\ell},\mathbf{s}_{f\ell}) = \mathcal{N}_c(\mathbf{x}_{f\ell};\mathbf{A}_{f\ell}\mathbf{s}_{f\ell},\var_f \mathbf{I}_I)$.

\subsection{The Channel Model}

A straightforward extension of~\cite{ozerov10mu} to time-varying linear filters is unfeasible. 
Indeed, instead of estimating the $I \times J \times F$ complex parameters of all $\mathbf{A}_{f}$, one would have to estimate the $I \times J \times F \times L$ complex parameters of all $\mathbf{A}_{f\ell}$ (with only $I \times F \times L$ observations).
In order to circumvent this issue, we model the mixing matrix $\mathbf{A}_{f\ell}$ as a latent variable and parameterize its temporal evolution, with much less parameters.

\begin{figure}[t]
\begin{center} 
\begin{tikzpicture}[scale=2,node distance=1.4cm,inner sep=0pt]
\node[rectangle,fill=none,draw,ultra thin, minimum width = 1.4cm,minimum height= .55cm]  (H)                          {$w_{fk},h_{k\ell}$};
\node[circle,fill=none,draw,minimum size = 1cm]                                          (S)   [right =.55cm of H]    {$\mathbf{s}_{f\ell}$};
\node[rectangle,fill=none,draw,ultra thin, minimum width = 1.2cm,minimum height= .65cm]  (muA) [above of = S     ]    {$\symvec{\mu}^a_f,\symmat{\Sigma}^a_f$};
\node[circle,fill=none,draw,minimum size = 1cm]                                          (A)   [right of = muA   ]    {$\mathbf{a}_{:,f\ell}$};
\node[double,circle,double distance=1pt,fill=metal,draw,thin, minimum size = 1.12cm]     (X)   [right of = S     ]    {\textcolor{white}{$\mathbf{x}_{f\ell}$}};
\node[rectangle,draw,ultra thin,minimum size =  .6cm]                                    (vX)  [right of = X     ]    {$\var_f$};
\path (S)   edge [thick,->] (X);
\path (A)   edge [thick,->] (X);
\path (muA) edge [thick,->] (A);
\path (H)   edge [thick,->] (S);
\path (vX)  edge [thick,->] (X);
\path (A) edge [loop right,looseness = 4.1,thick,->] node {$\;	\mathbf{a}_{:,f \ell-1}$} (A);
\end{tikzpicture}
\end{center}
\caption{ \label{fig:graphicalModel}Graphical model for time-varying convolutive mixtures with NMF source model. 
Latent variables are represented with circles, observations with double circles, deterministic parameters with rectangles, and temporal dependencies with self loops.}
\end{figure}
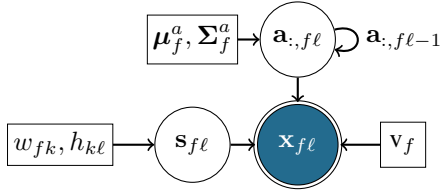

For this purpose, we first vectorize $\mathbf{A}_{f\ell}$ by vertically concatenating its $J$ columns $\{\mathbf{a}_{j,f\ell}\}_{j=1}^J$
into a single vector $\mathbf{a}_{:,f\ell} \in \mathbb{C}^{IJ}$,
i.e. $\mathbf{a}_{:,f\ell} = \text{vec}(\mathbf{A}_{f\ell}) = [\mathbf{a}_{1,f\ell}^\top \ldots \mathbf{a}_{J,f\ell}^\top]^\top$. 
In the following $\mathbf{a}_{:,f\ell}$ is referred to as the \emph{mixing vector}.
Then we assume that for every frequency $f$ the sequence of the $L$ unobserved mixing vectors $\{\mathbf{a}_{:,f\ell}\}_{\ell=1}^L$ is ruled by a first-order LDS,
where both the prior distribution and the process noise are assumed complex Gaussian. 
Formally, this writes:
\begin{align}
p(\mathbf{a}_{:,f\ell} | \mathbf{a}_{:,f\ell-1}) & = \mathcal{N}_c (\mathbf{a}_{:,f\ell};\mathbf{a}_{:,f\ell-1},\symmat{\Sigma}^a_f), \label{eq:temp1}\\
p(\mathbf{a}_{:,f1}) & = \mathcal{N}_c (\mathbf{a}_{:,f1};\symvec{\mu}^a_f,\symmat{\Sigma}^a_f),\label{eq:temp2}
\end{align}
where the mean vector $\symvec{\mu}^a_f \in \mathbb{C}^{IJ}$ and the \emph{evolution} covariance matrix $\symmat{\Sigma}^a_f \in \mathbb{C}^{IJ \times IJ}$ are parameters to be estimated.
$\symmat{\Sigma}^a_f$ is expected to reflect the amplitude of variations in the channel.
Importantly, the time-invariant mixing model of \cite{ozerov10mu} corresponds to the particular case in the proposed model when $\symmat{\Sigma}^a_f 
\rightarrow  \mathbf{0}_{IJ\times IJ}$. Indeed, in that case the latent state $\mathbf{a}_{:,f\ell}$ collapses to 
$\mathbf{a}_{:,f1}$ and hence the mixing matrix $\mathbf{A}_{f\ell}$ reduces to its time-invariant version $\mathbf{A}_f$. The complete graphical model of the proposed probabilistic model for audio source separation of time-varying convolutive mixtures is given in 
Fig.~\ref{fig:graphicalModel}.


The standard way to perform inference in LDS is the \textit{Kalman smoother} (or the \textit{Kalman filter} if only causal observations are used). 
Eq. \eqref{eq:mixing} defines the \emph{observation model} of the Kalman smoother.\footnote{The vectorized form of the latent mixing filters can be made explicit in the observation model by rewriting it as $\mathbf{x}_{f\ell} = \big(\mathbf{s}_{f\ell}^\top \otimes \mathbf{I}_I \big) \mathbf{a}_{:,f\ell} + \mathbf{b}_{f\ell}$, with $\otimes$ denoting the Kronecker matrix product.} However, since part of the observation model, for instance $\mathbf{s}_{f\ell}$, is a latent variable, the direct application of the classical Kalman technique is infeasible in our case. In other words, we need to infer both latent variables: the mixing filters and the sources/components.
For this purpose, in the next section we introduce a VEM procedure that alternates between (i) the complex Kalman smoother to infer the mixing filters sequence, (ii) the Wiener filter to estimate the sources and (iii) update rules for the parameters. Importantly, this result is a consequence of the joint effect of the proposed model and the variational approximation.

%

\section{VEM for Source Separation} \label{sec:varEM}

In this section, we present the proposed variational EM algorithm that alternates between the inference of the latent variables and the update of the parameters. 
We start with stating the principle of VEM. Then we present the E-step, farther decomposed in an E-A step for the mixing vector sequence and an E-S/C step for source/component coefficients, and then the M-step. The following notations are introduced: $\mathbb{E}_{q}$ is the expectation with respect to $q$,  $\hat{\mathbf{z}} = \mathbb{E}_{q(\mathbf{z})}[\mathbf{z}]$ is the posterior mean vector of a random vector $\mathbf{z}$, $\symmat{\Sigma}^{\eta z} = \mathbb{E}_{q(\mathbf{z})}[(\mathbf{z}-\hat{\mathbf{z}})
(\mathbf{z}-\hat{\mathbf{z}})^\hermitian]$ is its posterior covariance matrix, and $\mathbf{Q}^{\eta z} = \mathbb{E}_{q(\mathbf{z})}[\mathbf{z}{\mathbf{z}}^\hermitian] =  
\symmat{\Sigma}^{\eta z} + \hat{\mathbf{z}}\hat{\mathbf{z}}^\hermitian$ is its second-order posterior moment.
In general, superscript $\eta$ denotes parameters of \emph{posterior} distributions, whereas no superscript denotes parameters of \emph{prior} distributions. 
The posterior mean is the estimate of the corresponding latent variable, provided by our algorithm.
Also, let $\Sigma_{kg,f\ell}$ denote the $(k,g)$-th entry of matrix $\symmat{\Sigma}_{f\ell}$.
Let $\overset{ct}{=}$ denote equality up to an additive term that is independent of the variable at stake, and let $\text{tr}\{\cdot\}$ denote the trace operator.
For brevity $\mathbf{a}_{:,f1:L} = \{\mathbf{a}_{:,f\ell}\}_{\ell=1}^L$ denotes the whole sequence of mixing vectors at frequency $f$.

\subsection{Variational Inference Principle}

EM is a standard procedure to find maximum likelihood (ML) estimates in the presence of hidden variables \cite{McLachlanEM,bishop06pa}. 
By alternating between the evaluation of the posterior distribution of the hidden variables (E-step) and the maximization of the expected complete-data log-likelihood 
(M-step), EM provides ML parameter estimates from the set of observations $\{\mathbf{x}_{f\ell}\}_{f,\ell=1}^{F,L}$. In this work the set of 
hidden variables $\mathcal{H} = \{ \mathbf{a}_{:,f\ell},\mathbf{s}_{f\ell},\mathbf{c}_{f\ell} \}_{f,\ell=1}^{F,L}$ consists of the mixing vectors and the source
(or the component) coefficients.
The parameter set $\theta=\{\symvec{\mu}^a_f,\symmat{\Sigma}^a_f,w_{fk},h_{k\ell},\var_f\}_{f,\ell,k=1}^{F,L,K}$ 
consists of the channel evolution parameters, the source NMF parameters, and the variance of the sensor noise.

In our case, the posterior distribution of the latent variables, $q(\mathcal{H}) = p(\mathcal{H} | \{\mathbf{x}_{f\ell}\}_{f,\ell=1}^{F,L};\theta)$ 
cannot be expressed in closed-form. Therefore we develop a variational inference procedure \cite{bishop06pa},
\cite{smidlVAR}, based on the following principle. First, $q(\mathcal{H})$ is assumed to factorize into marginal posterior distributions over a partition of the latent variables. 
An approximation of the marginal posterior distribution of a subset of latent variables $\mathcal{H}_0 \subseteq \mathcal{H}$ is then computed with:
\begin{align} \label{eq:VBapprox}
q(\mathcal{H}_0) \propto \exp \left(\mathbb{E}_{q(\mathcal{H}/\mathcal{H}_0)} \left[ \log 
p(\mathcal{H},\{\mathbf{x}_{f\ell}\}_{f,\ell=1}^{F,L};\theta)  \right]\right),
\end{align}
where $q(\mathcal{H}/\mathcal{H}_0)$ is the \addnote[q-approx-1]{1}{approximation of the joint posterior distribution}~of
all hidden variables, except the subset $\mathcal{H}_0$. 
Subsequently, $q(\mathcal{H})$ can be inferred in an alternating manner for each 
$\mathcal{H}_0 \subset \mathcal{H}$. In the present work, we assume that the mixing filters and the source coefficients are conditionally independent given the observations. 
Therefore, the \addnote[q-approx-2]{1}{posterior distribution}\footnote{From now on, we abuse the language and refer to $q$ as the posterior 
distribution, even if technically it is only a variational approximation of it.} naturally factorizes as:
\begin{equation}
q(\mathcal{H}) \approx \prod \limits_{f=1}^F q(\mathbf{a}_{:,f1:L}) \prod \limits_{f,\ell=1}^{F,L} q(\mathbf{s}_{f\ell}). \label{eq:meanFieldApprox}
\end{equation}
Note that the factorization over frequency (for both sources and filters) and over time (for the sources) arises naturally from the prior 
\addnote[factorization]{1}{distributions and from the observation model~(\ref{eq:mixing})}.

\subsection{E-A Step} \label{sec:EAstep}

Using \eqref{eq:VBapprox} it is straightforward to show that the joint posterior distribution of the mixing vector sequence writes: 
\begin{equation}
\!q(\mathbf{a}_{:,f1:L}) \!\propto p(\mathbf{a}_{:,f1:L}) \! \prod_{\ell=1}^L \!
\exp \left( \mathbb{E}_{q(\mathbf{s}_{f\ell})} \big[ \log p(\mathbf{x}_{f\ell} | \mathbf{A}_{f\ell},\mathbf{s}_{f\ell}) \big]\right).\label{eq:filtersJointDistribution}
\end{equation}
We have:
\begin{align}
\nonumber &\mathbb{E}_{q(\mathbf{s}_{f\ell})} \big[ \log p(\mathbf{x}_{f\ell} | \mathbf{A}_{f\ell},\mathbf{s}_{f\ell}) \big] \overset{ct}{=} \nonumber \\
& \quad  -\text{tr} \bigg\{ \mathbb{E}_{q(\mathbf{s}_{f\ell})} \Big[ (\mathbf{x}_{f\ell}-\mathbf{A}_{f\ell}\mathbf{s}_{f\ell}) 
(\mathbf{x}_{f\ell}-\mathbf{A}_{f\ell}\mathbf{s}_{f\ell})^\hermitian \Big] {\var_f}^{-1} \bigg\} \overset{ct}{=} \nonumber \\
& \quad   -\text{tr}\bigg\{ 
\frac{\mathbf{I}_I}{\var_f}
\big( \mathbf{A}_{f\ell} - \mathbf{M}^{\iota a}_{f\ell} \big) \mathbf{Q}^{\eta s}_{f\ell} \big( \mathbf{A}_{f\ell} 
- \mathbf{M}^{\iota a}_{f\ell} \big)^\hermitian \bigg\}, \label{eq:EqS}
\end{align}
where $\mathbf{M}^{\iota a}_{f\ell} = \mathbf{x}_{f\ell} {\hat{\mathbf{s}}_{f\ell}}^\hermitian (\mathbf{Q}^{\eta s}_{f\ell})^{-1} \in \mathbb{C}^{I 
\times J}$, with $\hat{\mathbf{s}}_{f\ell}$ and $\mathbf{Q}^{\eta s}_{f\ell}$ provided by the E-S step in Section~\ref{sec:ESstep}.
By defining $\symmat{\mu}^{\iota a}_{f\ell} = \text{vec} 
(\mathbf{M}^{\iota a}_{f\ell})  \in \mathbb{C}^{IJ}$, \eqref{eq:EqS} can be reorganized as:
\begin{align}
\nonumber \mathbb{E}_{q(\mathbf{s}_{f\ell})} & \big[ \log p(\mathbf{x}_{f\ell} | \mathbf{A}_{f\ell},\mathbf{s}_{f\ell}) \big] \overset{ct}{=} \\
 - & (\mathbf{a}_{:,f\ell} - \symmat{\mu}^{\iota a}_{f\ell})^\hermitian
{\bigg( {\mathbf{Q}^{\eta s}_{f\ell}}^{\top} \otimes \frac{\mathbf{I}_I}{\var_f} \bigg)}
(\mathbf{a}_{:,f\ell} - \symmat{\mu}^{\iota a}_{f\ell}).  \label{eq:EqS2}
\end{align}
Let us define $\symmat{\Sigma}^{\iota a}_{f\ell} = \big({\mathbf{Q}^{\eta s}_{f\ell}}^{\top} \otimes \mathbf{I}_I\var_f^{-1}\big)^{-1} \in \mathbb{C}^{IJ \times IJ}$.
This matrix is Hermitian positive definite and (\ref{eq:EqS2}) characterizes a complex Gaussian distribution with mean $\symvec{\mu}^{\iota a}_{f\ell}$ and covariance $\symmat{\Sigma}^{\iota a}_{f\ell}$. 
By substituting \eqref{eq:EqS2} in \eqref{eq:filtersJointDistribution}, we obtain:
\begin{equation}
\!q(\mathbf{a}_{:,f1:L}) \!\propto p(\mathbf{a}_{:,f1:L}) \! \prod_{\ell=1}^L \!
\mathcal{N}_c(\symvec{\mu}^{\iota a}_{f\ell};\mathbf{a}_{:,f\ell},\symmat{\Sigma}^{\iota a}_{f\ell}).\!\!\!\label{eq:filtersJointDistribution2}
\end{equation}
Functional $\mathcal{N}_c(\symvec{\mu}^{\iota a}_{f\ell};\mathbf{a}_{:,f\ell},\symmat{\Sigma}^{\iota a}_{f\ell})$ can be viewed as an 
\emph{instantaneous} distribution of a \emph{measured} vector $\symvec{\mu}^{\iota a}_{f\ell}$, conditioned to the \emph{hidden} variable 
$\mathbf{a}_{:,f\ell}$.
Henceforth one recognizes that \eqref{eq:filtersJointDistribution2} represents an LDS with continuous hidden state variables $\{\mathbf{a}_{:,f\ell}\}_{\ell=1}^L$,
transition distribution given by \eqref{eq:temp1}, initial distribution given by \eqref{eq:temp2}, and \emph{emission} distribution given by
$\mathcal{N}_c(\symvec{\mu}^{\iota a}_{f\ell};\mathbf{a}_{:,f\ell},\symmat{\Sigma}^{\iota a}_{f\ell})$.
Subsequently the marginal posterior distribution of each hidden state, $q(\mathbf{a}_{:,f\ell})$, 
can be calculated recursively using a \emph{forward-backward} algorithm \cite{bishop06pa}, aka \emph{Kalman smoother}.

\subsubsection{Forward-backward algorithm}
Given the LDS parameters, a forward-backward algorithm computes an estimate $\hat{\mathbf{a}}_{:,f\ell}$ 
for all $\ell$ by taking into account all causal measurements (from $1$ to $\ell$) and anti-causal measurements (from $\ell+1$ to $L$). 
The implementation of the forward-backward algorithm thus consists of a recursive forward pass and a recursive backward pass. 
Different variants for this algorithm are available. The \emph{forward-backward} procedure that we specifically designed to infer \eqref{eq:filtersJointDistribution2} is described below. Because of the form of \eqref{eq:temp1}, all covariance updates of this forward-backward algorithm are computable using only additions and matrix inversion. Indeed it is desirable to avoid subtractions and matrix multiplications of covariance matrices since these operations do not guarantee that (with Hermitian operands) the resulting matrix is Hermitian. As a result, the proposed Kalman smoother was found to be very stable from a numerical point of view. In addition, since all distributions under consideration are complex Gaussian, the outcome of the forward-backward recursions will also be complex Gaussian \cite{bishop06pa}.


The forward pass recursively provides the joint distribution of the state variable and the causal observations. The mean vector $\symvec{\mu}^{\phi a}_{f\ell} \in \mathbb{C}^{IJ}$ and covariance matrix $\symmat{\Sigma}^{\phi a}_{f\ell} \in \mathbb{C}^{IJ \times IJ}$ of this distribution are calculated as:
\begin{align} \label{eq:fV}
\!\symmat{\Sigma}^{\phi a}_{f\ell} &= {\Big( {\symmat{\Sigma}^{\iota \vphantom{\zeta} a}_{f\ell}}^{-1} + 
{\big( \symmat{\Sigma}^{\phi a}_{f\ell-1} + \symmat{\Sigma}^a_f \big)}^{-1} \Big)}^{-1},\\
\!\symvec{\mu}^{\phi a}_{f\ell} &= \symmat{\Sigma}^{\phi a}_{f\ell} \Big(
{\symmat{\Sigma}^{\iota \vphantom{\zeta} a}_{f\ell}}^{-1}\symvec{\mu}^{\iota a}_{f\ell}
+
{\big(  \symmat{\Sigma}^{\phi a}_{f\ell-1} + \symmat{\Sigma}^a_f  \big)}^{-1} \symvec{\mu}^{\phi a}_{f\ell-1}
\Big). \label{eq:fA}
\end{align}
The backward pass recursively provides the distribution of the anti-causal observations given the current state. The mean vector $\symvec{\mu}^{\beta a}_{f\ell} \in \mathbb{C}^{IJ}$ and covariance matrix $\symmat{\Sigma}^{\beta a}_{f\ell} \in \mathbb{C}^{IJ \times IJ}$ of this distribution are calculated as:
\begin{align} \label{eq:zV}
\symmat{\Sigma}^{\zeta a}_{f\ell} &= {\Big( {\symmat{\Sigma}^{\iota \vphantom{\zeta} a}_{f\ell+1}}^{-1} + {\symmat{\Sigma}^{\beta a}_{f\ell+1}}^{-1} \Big)}^{-1},\\
\symmat{\Sigma}^{\beta a}_{f\ell} &= \symmat{\Sigma}^a_f + \symmat{\Sigma}^{\zeta a}_{f\ell}, \label{eq:bV} \\
\symvec{\mu}^{\beta a}_{f\ell} &= \symmat{\Sigma}^{\zeta a}_{f\ell} \Big( {\symmat{\Sigma}^{\iota \vphantom{\zeta} a}_{f\ell+1}}^{-1}\symvec{\mu}^{\iota a}_{f\ell+1} + 
{\symmat{\Sigma}^{\beta a}_{f\ell+1}}^{-1} \symvec{\mu}^{\beta a}_{f\ell+1} \Big), \label{eq:bA}
\end{align}
where $\symmat{\Sigma}^{\zeta a}_{f\ell} \in \mathbb{C}^{IJ \times IJ}$ is an intermediate matrix that enables to express the backward recursion without subtractions.

\subsubsection{Posterior estimate of the mixing vector}
Let us now calculate the \emph{smoothed} estimate $\hat{\mathbf{a}}_{:,f\ell}$.
By composing the forward and the backward estimates, the marginal (frame-wise) posterior distribution of $\mathbf{a}_{:,f\ell}$ writes \cite{bishop06pa}:
\begin{align}
q(\mathbf{a}_{:,f\ell}) = \mathcal{N}_c(\mathbf{a}_{:,f\ell}; \hat{\mathbf{a}}_{:,f\ell},\symmat{\Sigma}^{\eta a}_{f\ell}),
\end{align}
with $\symmat{\Sigma}^{\eta a}_{f\ell} \in \mathbb{C}^{IJ \times IJ}$ and $\hat{\mathbf{a}}_{:,f\ell} \in \mathbb{C}^{IJ}$ computed as:
\begin{align} \label{eq:Va}
\symmat{\Sigma}^{\eta a}_{f\ell } &= {\left( {\symmat{\Sigma}^{\phi a}_{f\ell}}^{-1} + {\symmat{\Sigma}^{\beta a}_{f\ell}}^{-1}\right)}^{-1}, \\
\hat{\mathbf{a}}_{:,f\ell} &= \symmat{\Sigma}^{\eta a}_{f\ell }
\left(
{\symmat{\Sigma}^{\phi a}_{f\ell}}^{-1} \symvec{\mu}^{\phi a}_{f\ell}
+
{\symmat{\Sigma}^{\beta a}_{f\ell}}^{-1} \symvec{\mu}^{\beta a}_{f\ell}
\right). \label{eq:A}
\end{align}

\subsubsection{Joint posterior distribution of a pair of successive mixing vectors}
This joint distribution will be needed to update $\symmat{\Sigma}^a_f$ in Section~\ref{sec:Mstep2}.
Let $\mathbf{a}_{:,f\{\ell+1,\ell\}} = {\big[\mathbf{a}_{:,f\ell+1}^\top,\mathbf{a}_{:,f\ell}^\top \big]}^\top \in \mathbb{C}^{2IJ}$ denote the joint variable.
By marginalizing out all mixing vectors except $\mathbf{a}_{:,f\ell+1},\mathbf{a}_{:,f\ell}$ in \eqref{eq:filtersJointDistribution2},
the joint posterior distribution $q(\mathbf{a}_{:,f\{\ell+1,\ell\}})$ can be identified to be also a Gaussian distribution with 
mean vector $\symvec{\mu}^{\xi a}_{f\ell} \in \mathbb{C}^{2IJ}$ and covariance matrix $\symmat{\Sigma}^{\xi a}_{f\ell} \in \mathbb{C}^{2IJ \times 2IJ}$ computed as:
\begin{align} \label{eq:jVa}
\symmat{\Sigma}^{\xi a}_{f\ell} &= \begin{bmatrix}
{\symmat{\Sigma}^{\zeta a}_{f\ell}}^{-1} + {\symmat{\Sigma}^{\vphantom{\zeta} a}_f}^{-1} & {-\symmat{\Sigma}^{\vphantom{\zeta} a}_f}^{-1} \\
{-\symmat{\Sigma}^{\vphantom{\zeta} a}_f}^{-1} & {\symmat{\Sigma}^{\phi a}_{f\ell}}^{-1} + {\symmat{\Sigma}^{\vphantom{\zeta} a}_f}^{-1}
\end{bmatrix}^{-1},\\
\symvec{\mu}^{\xi a}_{f\ell} &= \symmat{\Sigma}^{\xi a}_{f\ell}
\begin{bmatrix}
{\left(  {\symmat{\Sigma}^{\zeta a}_{f\ell}}^{-1} \symvec{\mu}^{\beta a}_{f\ell+1} \right)}^\top ,
{\left(  {\symmat{\Sigma}^{\phi a}_{f\ell}}^{-1} \symvec{\mu}^{\phi a}_{f\ell}     \right)}^\top 
\end{bmatrix}^\top. \label{eq:jA}
\end{align}
Note here the role of $\symmat{\Sigma}^{\zeta a}_{f\ell}$ that is to describe the uncertainty of $\symvec{\mu}^{\beta a}_{f\ell+1}$ but without incorporating the additional uncertainty of the transition variance $\symmat{\Sigma}^a_f$, as the transition from $\mathbf{a}_{:,f\ell}$ to $\mathbf{a}_{:,f\ell+1}$ is explicitly defined by the joint variable $\mathbf{a}_{:,f\{\ell+1,\ell\}}$.

\subsection{E-S Step and E-C Step}  \label{sec:ESstep}

From \eqref{eq:VBapprox}, the posterior distribution of the sources writes:
\begin{align} \label{eq:qS}
q(\mathbf{s}_{f\ell}) \propto p(\mathbf{s}_{f\ell})
\exp \left(\mathbb{E}_{q(\mathbf{a}_{:,f\ell})} \big[ \log p(\mathbf{x}_{f\ell} | \mathbf{A}_{f\ell},\mathbf{s}_{f\ell}) \big] \right).
\end{align}
Using \addnote[expectation-corr]{1}{(\ref{eq:mixing})}, the expectation in (\ref{eq:qS}) computes:
 \begin{align}
 &\mathbb{E}_{q(\mathbf{a}_{:,f\ell})} \big[ \log p(\mathbf{x}_{f\ell} | \mathbf{A}_{f\ell},\mathbf{s}_{f\ell}) \big] 
 \overset{ct}{=} \nonumber \\
& \frac{1}{\var_f}\text{tr}\Big\{
 \mathbf{s}_{f\ell} \big( \hat{\mathbf{A}}_{f\ell}^\hermitian \mathbf{x}_{f\ell} \big)^\hermitian 
+ \big( \hat{\mathbf{A}}_{f\ell}^\hermitian \mathbf{x}_{f\ell}\big)  \mathbf{s}_{f\ell}^\hermitian 
-  \mathbf{U}_{f\ell} \mathbf{s}_{f\ell} \mathbf{s}_{f\ell}^\hermitian  \Big\} , \label{eq:EqA}
 \end{align} 
 where $\hat{\mathbf{A}}_{f\ell} = \mathbb{E}_{q(\mathbf{a}_{:,f\ell})}[\mathbf{A}_{f\ell}] \in \mathbb{C}^{I \times J}$ is a matrix constructed 
 from $\hat{\mathbf{a}}_{:,f\ell}$ (i.e. the reverse operation of column-wise vectorization), 
and $\mathbf{U}_{f\ell} = \mathbb{E}_{q(\mathbf{a}_{:,f\ell})} [\mathbf{A}_{f\ell}^\hermitian \mathbf{A}_{f\ell}] \in \mathbb{C}^{J \times J}$.
Of course, $\mathbf{U}_{f\ell}$ is closely related to $\mathbf{Q}_{f\ell}^{\eta a}$. Indeed, if we define $\mathbf{Q}^{\eta a}_{jr,f\ell} = 
\mathbb{E}_{q(\mathbf{a}_{:,f\ell})}[ \mathbf{a}_{j,f\ell} \mathbf{a}_{r,f\ell}^\hermitian ]$ as the $(j,r)$-th $I \times I$ subblock of 
$\mathbf{Q}_{f\ell}^{\eta a}$, then each entry $U_{jr,f\ell}$ of $\mathbf{U}_{f\ell}$ is simply given by:
\begin{align}
U_{jr,f\ell} = \mathbb{E}_{q(\mathbf{a}_{:,f\ell})}[ \mathbf{a}_{j,f\ell}^\hermitian \mathbf{a}_{r,f\ell} ] = \text{tr}\Big\{ \mathbf{Q}^{\eta 
a}_{rj,f\ell} \Big\}. \label{eq:Ut}
\end{align}
Eq. (\ref{eq:EqA}) is an incomplete quadratic form in $\mathbf{s}_{f\ell}$. 
Combining in \eqref{eq:qS} this quadratic form with the quadratic form of the source prior $p(\mathbf{s}_{f\ell})$, we obtain a multivariate Gaussian:
\begin{equation} \label{eq:qSunexpanded}
q(\mathbf{s}_{f\ell}) = \mathcal{N}_c(\mathbf{s}_{f\ell};\hat{\mathbf{s}}_{f\ell},\symmat{\Sigma}^{\eta s}_{f\ell}),
\end{equation}
with mean vector $\hat{\mathbf{s}}_{f\ell} \in \mathbb{C}^J$ and covariance matrix $\symmat{\Sigma}^{\eta s}_{f\ell} \in \mathbb{C}^{J \times J}$ given by:
\begin{align} \label{eq:Vs}
\symmat{\Sigma}^{\eta s}_{f\ell} &= {\left[ \text{diag}_J 
\bigg( 
\frac{1}{\scalebox{.9}{$\sum \limits_{k \in \mathcal{K}_j}$}  w_{fk}h_{k\ell}}
\bigg) + \frac{\mathbf{U}_{f\ell}}{\var_f}\right]}^{-1},\\[1.5ex]
\hat{\mathbf{s}}_{f\ell} &= \symmat{\Sigma}^{\eta s}_{f\ell} \hat{\mathbf{A}}_{f\ell}^\hermitian \frac{\mathbf{x}_{f\ell}}{\var_f}. \label{eq:S}
\end{align}
Remarkably, (\ref{eq:S}) 
has a form similar to the source estimator in \cite{ozerov10mu}, namely a Wiener filtering estimator, with two notable differences. First, in \cite{ozerov10mu} the mixing matrix 
is an estimated parameter, whereas here it is the posterior expectation $\hat{\mathbf{A}}_{f\ell}$ of the latent mixing matrix.
Second, the source posterior precision matrix $(\symmat{\Sigma}^{\eta s}_{f\ell})^{-1}$ is built by summation of (i)~the sensor precision $1/\var_f$ 
distributed over the sources with the unit-less quantity $\mathbf{U}_{f\ell}$, and of  (ii)~the diagonal prior precision of the source coefficients given 
by the NMF model (as in \cite{ozerov10mu}). In other words, the a posteriori uncertainty of the sources encompasses the a priori uncertainty (the 
NMF), the channel noise ($\var_f$), and the channel uncertainty ($\mathbf{U}_{f\ell}$).

A similar E-step can be applied to the source components $\mathbf{c}_{f\ell}$. This will be used Section~\ref{sec:NMF} to optimize the NMF 
parameters. For this aim, we simply replace $\mathbf{A}_{f\ell}$ with $\mathbf{A}_{f\ell}\mathbf{G}$, and $p(\mathbf{s}_{f\ell})$ with 
$p(\mathbf{c}_{f\ell})$, obtaining again a complex Gaussian for the posterior distribution of the components:
\begin{equation}
q(\mathbf{c}_{f\ell}) = \mathcal{N}_c(\mathbf{c}_{f\ell};\hat{\mathbf{c}}_{f\ell},\symmat{\Sigma}^{\eta c}_{f\ell}),
\end{equation}
with parameters $\hat{\mathbf{c}}_{f\ell} \in \mathbb{C}^K$ and $\symmat{\Sigma}^{\eta c}_{f\ell} \in \mathbb{C}^{K \times K}$ given by:
\begin{align}
\symmat{\Sigma}^{\eta c}_{f\ell} & = {\left[ \text{diag}_K \bigg( \frac{1}{w_{fk}h_{k\ell}}\bigg) + \mathbf{G}^\top \frac{\mathbf{U}_{f\ell}}{\var_f} \mathbf{G} \right]}^{-1}, \label{eq:SigmaC} \\
\hat{\mathbf{c}}_{f\ell} & = \symmat{\Sigma}^{\eta c}_{f\ell} \mathbf{G}^\top \hat{\mathbf{A}}_{f\ell}^\hermitian \frac{\mathbf{x}_{f\ell}}{\var_f}. \label{eq:C}
\end{align}
Again, (\ref{eq:C}) is a Wiener filtering estimator, here at the source component vector level. Note that left-multiplication of both sides of (\ref{eq:C}) by $\mathbf{G}$ naturally leads to (\ref{eq:S}).  

\subsection{Outline of the Maximization Step}

Once we have the posterior distributions of the variables in $\mathcal{H}$, the \emph{expected complete-data log-likelihood} 
$\mathcal{L}(\theta) = \mathbb{E}_{q(\mathcal{H})} \log p \big( \mathcal{H},\{\mathbf{x}_{f\ell}\}_{f,\ell=1}^{F,L};\theta \big)$
is maximized with respect to the parameters.
The analytic expression of $\mathcal{L}(\theta)$ is
\begin{align}
\mathcal{L}(\theta) & = \sum \limits_{f,\ell=1}^{F,L} 
\mathbb{E}_{q(\mathbf{a}_{:,f\ell})q(\mathbf{s}_{f\ell})}  \big[\log \mathcal{N}_c(\mathbf{x}_{f\ell};\mathbf{A}_{f\ell}\mathbf{s}_{f\ell},\var_f \mathbf{I}_I) \big] \nonumber \\
& + \sum \limits_{f,\ell=1}^{F,L}
\mathbb{E}_{q(\mathbf{c}_{f\ell})} \big[\log \mathcal{N}_c\left(\mathbf{c}_{f\ell};\mathbf{0},\text{diag}_K(w_{fk}h_{k\ell})\right)\big]  \nonumber \\
& + \sum \limits_{f=1}^F \bigg( \sum \limits_{\ell=1}^{L-1} 
\mathbb{E}_{q(\mathbf{a}_{:,f\{\ell+1,\ell\}})} \big[ \log 
\mathcal{N}_c\left(\mathbf{a}_{:,f\ell+1}; \mathbf{a}_{:,f\ell},\symmat{\Sigma}^a_f\right)
 \big]  \nonumber \\
& + \mathbb{E}_{q(\mathbf{a}_{:,f1})} \big[ \log \mathcal{N}_c\left(\mathbf{a}_{:,f1}; \symvec{\mu}^a_f,\symmat{\Sigma}^a_f\right) \big] \bigg). \label{eq:ECDLL}
\end{align}
Notice that \eqref{eq:ECDLL} can be optimized w.r.t. the microphone noise parameters, the channel parameters, or the NMF parameters, independently.

\subsection{M-V Step}
Derivating $\mathcal{L}(\theta)$ w.r.t. $\var_f$, and setting the result to zero, leads to the following update:
\begin{align} \label{eq:Mx}
\var_f & = \frac{1}{LI} \sum \limits_{\ell=1}^L \Big( \mathbf{x}_{f\ell}^\hermitian \mathbf{x}_{f\ell} - \mathbf{x}_{f\ell}^\hermitian
\hat{\mathbf{A}}_{f\ell}\hat{\mathbf{s}}_{f\ell}  \nonumber \\
& - \big( \hat{\mathbf{A}}_{f\ell} \hat{\mathbf{s}}_{f\ell} \big)^\hermitian \mathbf{x}_{f\ell} +
\text{tr}\big\{\mathbf{U}_{f\ell} \mathbf{Q}^{\eta s}_{f\ell} \big\} \Big),
\end{align}
which resembles the estimator obtained in \cite{ozerov10mu}.

\subsection{M-A Step}
\label{sec:Mstep2}
Optimizing $\mathcal{L}(\theta)$ w.r.t. the prior mean $\symvec{\mu}^a_f$ results in the following update:
\begin{equation} \label{eq:muA}
\symvec{\mu}^a_f = \hat{\mathbf{a}}_{f1}.
\end{equation}
The ML initial vector is thus the posterior mean vector for $\ell=1$. The way the E-A step was designed, \eqref{eq:muA} becomes rather important.

As for $\symmat{\Sigma}^a_f$, the terms of $\mathcal{L}(\theta)$ that depend on this parameter reduce to:
\begin{align}
\mathcal{L}(\symmat{\Sigma}^a_f)  \equiv & \sum \limits_{\ell=1}^{L-1} \mathbb{E}_{q(\mathbf{a}_{:,f\{\ell+1,\ell\}})} 
\left[\log \mathcal{N}_c\left(\mathbf{a}_{:,f\ell+1}; \mathbf{a}_{:,f\ell},\symmat{\Sigma}^a_f\right)\right]  \nonumber \\
& + \mathbb{E}_{q(\mathbf{a}_{:,f1})} \left[\log \mathcal{N}_c\left(\mathbf{a}_{:,f1}; \symvec{\mu}^a_f,\symmat{\Sigma}^a_f\right)\right] \nonumber \\
\overset{ct}{=} &
- \text{tr}\Big\{  {\symmat{\Sigma}^{\vphantom{\zeta} a}_f}^{-1} \symmat{\Sigma}^{\eta a}_{f1} \Big\} \nonumber \\
& - \text{tr}\left\{ 
\begin{bmatrix}
{\symmat{\Sigma}^{\vphantom{\zeta} a}_f}^{-1} & - {\symmat{\Sigma}^{\vphantom{\zeta} a}_f}^{-1} \\
- {\symmat{\Sigma}^{\vphantom{\zeta} a}_f}^{-1} & {\symmat{\Sigma}^{\vphantom{\zeta} a}_f}^{-1}
\end{bmatrix} \mathbf{Q}^{\xi a}_f
\right\}
-L \log|\symmat{\Sigma}^a_f|  \nonumber \\
 = & -L \log|\symmat{\Sigma}^a_f| \nonumber \\
- &  \text{tr}\bigg\{ {\symmat{\Sigma}^{\vphantom{\zeta} a}_f}^{-1} \Big[
\symmat{\Sigma}^{\eta a}_{f1} + \mathbf{Q}^{\xi a}_{11,f} - \mathbf{Q}^{\xi a}_{12,f} - \mathbf{Q}^{\xi a}_{21,f} + \mathbf{Q}^{\xi a}_{22,f}
\Big] \bigg\}. \label{eq:objeV}
\end{align}
In the above equation $\mathbf{Q}^{\xi a}_f \in \mathbb{C}^{2IJ \times 2IJ}$ is the cumulate second-order joint posterior moment 
of $\mathbf{a}_{:,f\{\ell+1,\ell\}}$, and the four $\mathbf{Q}^{\xi a}_{nm,f}$ matrices are its $IJ \times IJ$ non-overlapping principal subblocks, i.e.:
\begin{align} \label{eq:jQa}
\mathbf{Q}^{\xi a}_f = \sum \limits_{\ell=1}^{L-1} \left( \symmat{\Sigma}^{\xi a}_{f\ell} + 
\symvec{\mu}^{\xi a}_{f\ell} {(\symvec{\mu}^{\xi a}_{f\ell})}^\hermitian \right) =
\begin{bmatrix}
\mathbf{Q}^{\xi a}_{11,f} & \mathbf{Q}^{\xi a}_{12,f} \\[1.05ex]
\mathbf{Q}^{\xi a}_{21,f} & \mathbf{Q}^{\xi a}_{22,f} 
\end{bmatrix}.
\end{align}
Derivating \eqref{eq:objeV} w.r.t. the entries of $\symmat{\Sigma}^a_f$, and setting the result to zero, yields \cite{hjorungnes07co}:
\begin{align}
\symmat{\Sigma}^a_f = \frac{1}{L}\left( 
\mathbf{Q}^{\xi a}_{11,f} - 
\mathbf{Q}^{\xi a}_{12,f} - 
\mathbf{Q}^{\xi a}_{21,f} + 
\mathbf{Q}^{\xi a}_{22,f} + \symmat{\Sigma}^{\eta a}_{f1} \right). \label{eq:eV} 
\end{align}

\subsection{M-C Step and M-S Step}
\label{sec:NMF}

The joint optimization of $\mathcal{L}(\theta)$ over $w_{fk}$ and $h_{k\ell}$ is non-convex. 
However alternate maximization is a classical solution to solve for a locally-optimal set of NMF parameters \cite{fevotte09no}. 
Calculating the derivatives of $\mathcal{L}(\theta)$ w.r.t. to $w_{fk}$ and $h_{k\ell}$ and setting the result to zero leads to the following update formulae:
%
\begin{align} \label{eq:nmf}
w_{fk} = \frac{1}{L} \sum \limits_{\ell=1}^L
\frac{ Q^{\eta c}_{kk,f\ell} }{h_{k\ell}},\quad
h_{k\ell} = \frac{1}{F} \sum \limits_{f=1}^F
\frac{ Q^{\eta c}_{kk,f\ell} }{w_{fk}}.
\end{align}
\addnote[nmf-iterations]{1}{This formulae can be iteratively applied until convergence,
although in an effort to avoid local optima, each of $w_{fk}$,  $h_{k\ell}$ was updated only once at each VEM iteration.}


\subsection{Estimation of Source Images} \label{sec:imageEstimation}
As is often the case in source separation, the proposed framework suffers from the well-known scale ambiguity, namely
the source signals and the mixing matrices can only be estimated up to (frequency-dependent) compensating multiplicative factors~\cite{comon10ha}.
To alleviate this problem and to be able to assess the performance of source separation, we consider the separation of the source images, i.e. the source signals as recorded by the microphones \cite{duong10un, sturmel12li}, instead of the (monophonic) source signals.
For this purpose, the inverse STFT is applied to $\{ \mathbb{E}_{q(\mathbf{a}_{:,f\ell},\mathbf{s}_{f\ell})}[\mathbf{a}_{j,f \ell}s_{j,f \ell}] = 
\hat{\mathbf{a}}_{j,f\ell}\hat{s}_{j,f\ell}\}_{f,\ell=1}^{F,L}$, where $\hat{\mathbf{a}}_{j,f\ell}$ is the $j$-th column of $\hat{\mathbf{A}}_{f\ell}$. 
The complete VEM separating $J$ sound sources from an $I$-channel time-varying mixture is outlined in Algorithm~\ref{eq:algo} (omitting STFT and inverse STFT for clarity).


\begin{spacing}{1}

\begin{algorithm}[t]
\caption{Proposed VEM for the separation of sound sources mixed with time-varying filters}
\label{eq:algo}
\begin{algorithmic}
\STATE \textbf{input} $\{\mathbf{x}_{f\ell}\}_{f,\ell=1}^{F,L}$, partition matrix $\mathbf{G}$, initial parameters $\theta$.
\vspace{0.1cm}
\STATE \textbf{initialize} posterior statistics $\hat{\mathbf{a}}_{:,f\ell},\symmat{\Sigma}^{\eta a}_{f\ell}$.
\vspace{0.05cm}
\REPEAT
\vspace{0.1cm}
\STATE \textbf{Variational E-step}
\vspace{0.1cm}
\STATE Calculate $\mathbf{Q}^{\eta a}_{f\ell} = \symmat{\Sigma}^{\eta a}_{f\ell} + 
\hat{\mathbf{a}}_{:,f\ell}\hat{\mathbf{a}}_{:,f\ell}^\hermitian$
and $\mathbf{U}_{f\ell}$ with \eqref{eq:Ut}.
\vspace{0.1cm}
\STATE \emph{E-S step:}
Compute $\symmat{\Sigma}^{\eta s}_{f\ell}$ with \eqref{eq:Vs} and $\hat{\mathbf{s}}_{f\ell}$ with \eqref{eq:S}.\\
\hspace{1.3cm} Then compute $\mathbf{Q}^{\eta s}_{f\ell} = \symmat{\Sigma}^{\eta s}_{f\ell}+\hat{\mathbf{s}}_{f\ell} \hat{\mathbf{s}}_{f\ell}^\hermitian$.
\vspace{0.1cm}
\STATE \emph{E-C step:} Compute $\Sigma^{\eta c}_{kk,f\ell}$ with \eqref{eq:Vc3} and $\hat{c}_{k,f\ell}$ with \eqref{eq:C2}.\\
\hspace{1.3cm} Then compute $Q^{\eta c}_{kk,f\ell} = \Sigma^{\eta c}_{kk,f\ell} + |\hat{c}_{k,f\ell}|^2$.
\vspace{0.1cm}
\STATE \emph{E-A step (Instantaneous Quantities):}\\
\hspace{1.3cm} Compute $({\symmat{\Sigma}^{\iota \vphantom{\zeta} a}_{f\ell}}^{-1} \symvec{\mu}^{\iota a}_{f\ell})$ with \eqref{eq:iotaA}.\\
\hspace{1.3cm} Compute ${\symmat{\Sigma}^{\iota \vphantom{\zeta} a}_{f\ell}}^{-1} = {\mathbf{Q}^{\eta s}_{f\ell}}^{\top} \otimes \mathbf{I}_I\var_f^{-1}$.
\vspace{0.1cm}
\STATE \emph{E-A step (Forward Pass):}\\
\hspace{1.3cm} Initialize $\symmat{\Sigma}^{\phi a}_{f1} = \big( {\symmat{\Sigma}^{\iota \vphantom{\zeta} a}_{f1}}^{-1} + {\symmat{\Sigma}^{\vphantom{\zeta} a}_f}^{-1} \big)^{-1}$. \\
\hspace{1.3cm} Initialize $\symvec{\mu}^{\phi a}_{f1} = \symmat{\Sigma}^{\phi a}_{f1} \big({\symmat{\Sigma}^{\iota \vphantom{\zeta} a}_{f1}}^{-1}\symvec{\mu}^{\iota a}_{f1} + 
{\symmat{\Sigma}^{\vphantom{\zeta} a}_f}^{-1} \symvec{\mu}^a_f \big)$.\\
\hspace{1.3cm}\textbf{for} $\ell : 2$ to $L$\\
\STATE \hspace{1.8cm}Compute $\symmat{\Sigma}^{\phi a}_{f\ell}$ with \eqref{eq:fV}, then $\symvec{\mu}^{\phi a}_{f\ell}$ with \eqref{eq:fA}.\\
\hspace{1.3cm}\textbf{end}\\
\vspace{0.1cm}
\STATE \emph{E-A step (Backward Pass):} \\
\hspace{1.3cm} Initialize $\symmat{\Sigma}^{\beta a}_{fL} = \symmat{\Sigma}^{\phi a}_{fL}$ and $\symvec{\mu}^{\beta a}_{fL} = \symvec{\mu}^{\phi a}_{fL}$.\\
\vspace{0.1cm}
\hspace{1.3cm}\textbf{for} $\ell : L-1$ to $1$\\
\STATE \hspace{1.8cm} Compute $\symmat{\Sigma}^{\zeta a}_{f\ell}$ with \eqref{eq:zV}. \\
\hspace{1.8cm} Then compute $\symmat{\Sigma}^{\beta a}_{f\ell}$ with \eqref{eq:bV}. \\ 
\hspace{1.8cm}  Then compute $\symvec{\mu}^{\beta a}_{f\ell}$ with \eqref{eq:bA}.\\
\hspace{1.3cm}\textbf{end}\\
\vspace{0.1cm}
\STATE \emph{E-A step (Posterior Marginal Statistics):}\\
\hspace{1.3cm} Compute $\symmat{\Sigma}^{\eta a}_{f\ell}$ with \eqref{eq:Va}.\\
\hspace{1.3cm}  Then compute $\hat{\mathbf{a}}_{:,f\ell}$ with \eqref{eq:A}.
\vspace{0.1cm}
\STATE \emph{E-A step (Pairwise Joint Posterior):} \\
\hspace{1.3cm} Compute $\symmat{\Sigma}^{\xi a}_{f\ell}$ with \eqref{eq:jVa}. \\
\hspace{1.3cm}  Then compute $\symvec{\mu}^{\xi a}_{f\ell}$ with \eqref{eq:jA}.\\
\hspace{1.3cm} Then compute $\mathbf{Q}^{\xi a}_f$ with \eqref{eq:jQa}.
\STATE \textbf{M-step}
\vspace{0.1cm}
\STATE \emph{M-v step:} Update $\var_f$ with \eqref{eq:Mx}.
\vspace{0.1cm}
\STATE \emph{M-A step:} Update $\symvec{\mu}^a_f$ with \eqref{eq:muA}, update $\symmat{\Sigma}^a_f$ with \eqref{eq:eV}. 
\vspace{0.1cm}
\STATE \emph{M-C step:} Alternately update $w_{fk}$ and $h_{k\ell}$ with \eqref{eq:nmf}.
\vspace{0.1cm}
\UNTIL convergence \\
\vspace{0.1cm}
\RETURN the estimated source images $\hat{\mathbf{a}}_{j,f\ell}\hat{s}_{j,f\ell}, j \in [1,J]$.
\end{algorithmic}
\end{algorithm}

\end{spacing}

\section{Implementation Issues} \label{sec:Implementation}

In this section we present some simplifications that our algorithm admits,
we give physical interpretations, and we discuss some numerical stability issues.

\subsubsection{Simplifying the LDS measurement vector}
The forward-backward procedure requires the quantity $({\symmat{\Sigma}^{\iota a}_{f\ell}}^{-1}\symvec{\mu}^{\iota a}_{f\ell})$, 
appearing in \eqref{eq:fA} and \eqref{eq:bA}. This can be computed as:
\begin{equation} \label{eq:iotaA}
{\symmat{\Sigma}^{\iota \vphantom{\zeta} a}_{f\ell}}^{-1}\symvec{\mu}^{\iota a}_{f\ell}  =\frac{1}{\var_f}  \text{vec} \big( \mathbf{x}_{f\ell} \hat{\mathbf{s}}_{f\ell}^\hermitian \big),
\end{equation}
thus sparing the inversion of $\symmat{\Sigma}^{\iota a}_{f\ell}$.

\subsubsection{Initializing the forward and backward recursions}
The forward-backward algorithm needs to set $\symmat{\Sigma}_{f1}^{\phi a}$ and $\symvec{\mu}^{\phi a}_{f1}$ for 
the first frame, and to set $\symmat{\Sigma}_{fL}^{\beta a}$ and $\symvec{\mu}^{\beta a}_{fL}$ for the last frame.
We observed faster convergence with the following choice.
At each VEM iteration, we set $\symmat{\Sigma}^{\phi a}_{f1} = \left( {\symmat{\Sigma}^{\iota \vphantom{\zeta} a}_{f1}}^{-1} + {\symmat{\Sigma}^{\vphantom{\zeta} a}_f}^{-1} \right)^{-1}$ and 
$\symvec{\mu}^{\phi a}_{f1} = \symmat{\Sigma}^{\phi a}_{f1} \left({\symmat{\Sigma}^{\iota \vphantom{\zeta} a}_{f1}}^{-1}\symvec{\mu}^{\iota a}_{f1} + 
{\symmat{\Sigma}^{\vphantom{\zeta} a}_f}^{-1} \symvec{\mu}^a_f \right)$. Then, we run the forward pass first. After it is completed we set $\symmat{\Sigma}^{\beta a}_{fL} 
= \symmat{\Sigma}^{\phi a}_{fL},~ \symvec{\mu}^{\beta a}_{fL} = \symvec{\mu}^{\phi a}_{fL}$, to initialize the backward pass.

\subsubsection{Avoiding $K \times K$ matrix construction} 
Eq. (\ref{eq:SigmaC}) is computationally demanding as it requires the construction of a $K \times K$ matrix (recall that $K \gg J$). 
Yet, it has been shown in Section \ref{sec:NMF} that one needs only the diagonal entries of $\mathbf{Q}^{\eta c}_{f\ell}$. 
Therefore we derive an alternative expression for $\Sigma^{\eta c}_{kk,f\ell}$ and $\hat{c}_{k,f\ell}$ that builds on the already computed
$\symmat{\Sigma}^{\eta s}_{f\ell}$ and $\hat{\mathbf{s}}_{f\ell}$ (which use operations only on $J \times J$ arrays).
Applying the \emph{Woodbury} identity to \eqref{eq:SigmaC} and some algebraic manipulations, one obtains:
\begin{align} \label{eq:Vc3}
\Sigma^{\eta c}_{kk,f\ell} = w_{fk}h_{k\ell} \left(1 - \frac{ w_{fk}h_{k\ell} 
\Big[ \mathbf{U}_{f\ell} \symmat{\Sigma}^{\eta s}_{f\ell} \Big]_{j_kj_k} }{ \var_f \sum \limits_{\rho \in \mathcal{K}_{j_k}} w_{f\rho}h_{\rho \ell} }
\right),
\end{align}
where $j_k$ is the index of the source that the $k^\text{th}$ component belongs to, and $[\cdot]_{j_kj_k}$ is the $j_k^\text{th}$ diagonal element of the $J \times J$ matrix 
in brackets. 
Additionally, $\hat{c}_{k,f\ell}$ can be expressed in a very simple way, independently of $\symmat{\Sigma}^{\eta c}_{f\ell}$:
\begin{align} \label{eq:C2}
\hat{c}_{k,f\ell} = w_{fk}h_{k\ell} \bigg[ \hat{\mathbf{A}}_{f\ell}^\hermitian \frac{\mathbf{x}_{f\ell}}{\var_f} 
- \mathbf{U}_{f\ell} \frac{\hat{\mathbf{s}}_{f\ell}}{\var_f} \bigg]_{j_k},
\end{align}
where $[\cdot]_{j_k}$ is the $j_k^\text{th}$ element of the $J \times 1$ vector in brackets.
Interestingly, \eqref{eq:C2} shows that $\hat{c}_{k,f\ell}$ is some kind of inpainting onto the mixture signal, whose
purpose is to equalize the filtered mixture with the sources.
Besides, \eqref{eq:Vc3} makes clear that if the value of $\var_f$ is high enough,
the posterior variance of $c_{k,f\ell}$ remains close to its prior value $w_{fk}h_{k\ell}$.
This justifies the use of a high initial value for $\var_f$ in cases where the NMF parameters are quite correctly initialized. 

\subsubsection{Ordering the steps} When building a (V)EM algorithm, 
the question of ordering the steps execution arises. Like the majority of EMs, 
our algorithm is sensitive to initialization (discussed in Section~\ref{sec:init}). We observed in practice that our algorithm is much more sensitive to the initialization of the NMF parameters $w_{fk},h_{k\ell}$ 
than to the initialization of (the posterior parameters of) the mixing vectors: $\symmat{\Sigma}^{\eta a}_{f\ell},\hat{\mathbf{a}}_{:,f\ell}$.
Therefore we choose to first infer the source/component statistics by running E-S/C and then infer the sequence of mixing vectors by running E-A.
As for the M-steps, they are independent and so they can be executed in any order after the E-steps.

\subsubsection{NMF scaling} 
When estimating the NMF parameters using \eqref{eq:nmf}, an arbitrary scale can circulate between $w_{fk},h_{k\ell}$ 
of a component \cite{ozerov10mu}.
Therefore one can consider scaling one of the factors, e.g. $w_{fk} \leftarrow w_{fk}/\sum_n w_{nk}$, 
so to have unit $L1$-norm vectors, and reciprocally scaling the other factors, e.g. $h_{k\ell} \leftarrow h_{k\ell} \sum_f w_{fk}$ for compensation.

%
%
%
\subsubsection{Numerical stability} 
We enforce matrices $\mathbf{U}_{f\ell}$ and $\symmat{\Sigma}^a_f$ to be Hermitian with $\symmat{\Sigma}^a_f \leftarrow 
\frac{1}{2}(\symmat{\Sigma}^a_f+{\symmat{\Sigma}^a_f}^\hermitian)$. \addnote[matrix-reg]{1}{We also regularized the updates of $\var_f$ and of 
$\symmat{\Sigma}^{\xi a}_{f\ell}$, by adding $10^{-7}$ and $10^{-7} \mathbf{I}_{2IJ}$ respectively.}
%

\subsubsection{Computational complexity}
Counting only matrix multiplications, inversions and the solution of linear systems (assuming cubic complexity) the complexity order of the proposed VEM algorithm is 
$\mathcal{O}\big(16FL(IJ)^3 + 5FLG + F(L-1)(2IJ)^3\big)$. The experiments of this paper were conducted with a \emph{HP Z800} desktop 4-core computer
(8 threads) \emph{Xeon E5620} CPU at $2.4$~\!GHz and $17.6$~\!GB of RAM. To process a $2$s $16$KHz stereo mixture, with $J=3$, $K=75$, $F=512$, 
$L=128$ our non-optimized implementation needs $30$s per iteration, running in \emph{MATLAB R2014a}, on \emph{Fedora 20}. 
\addnote[comp-block-wise]{1}{On the same data,
the block-wise adaptation of the baseline method requires $4$s for a complete iteration (an iteration for all blocks of frames). 
Hence, with this set-up, the complexity of the proposed method is about $8$ times larger than the complexity of the 
baseline method}. 

\section{Experimental Study} \label{sec:exp}

To assess the performance of the proposed model and associated VEM algorithm, we conducted a series of experiments with 2-channel time-varying convolutive mixtures of speech signals. Initialization is known to be a crucial step for the performance of (V)EM algorithms. In a general manner, EM-like algorithms have severe difficulties to converge to a relevant solution in totally blind setups (i.e. random initialization). 
A first series of experiments was thus conducted with simulated mixtures and artificially controlled (semi-blind) initialization of the VEM in order 
to extensively investigate its performance independently of initialization problems. Then a second series was conducted using a state-of-the art blind 
source separation method based on binary masking for the initialization. This latter configuration was first applied on simulated mixtures and then 
real-world recorded mixtures.

\begin{figure}[t]
\centering
\centerline{\includegraphics[width=0.45\columnwidth]{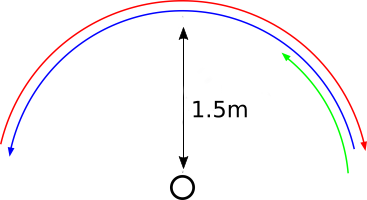} \hspace{1cm}
\includegraphics[width=0.45\columnwidth]{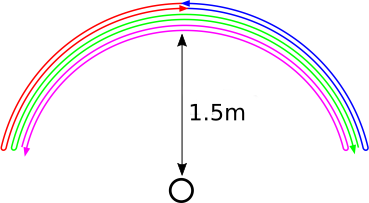}}
\caption{Type I (left) and II (right) source trajectories for the experiments with semi-blind initialization. In Type I, Sources $s_1$ (red) and $s_2$ (blue) move from $-\vartheta$ to $\vartheta$ and from $\vartheta$ to $-\vartheta$ respectively, while Source $s_3$ moves from $85^\circ$ to $45^\circ$. In Type II, sources move: from $0^\circ$ to $-\vartheta$ and back ($s_1$, red), from $0^\circ$ to $\vartheta$ and back ($s_2$, blue), from $-\vartheta$ to $\vartheta$ and back ($s_3$, purple) and from $\vartheta$ to $-\vartheta$ and back ($s_4$, green); note that $s_3$ and $s_4$ move twice as fast as $s_1$ and $s_2$. In this example, $\vartheta=75^\circ$.}
\label{fig:trajectories}
\end{figure}


\subsection{Experiments with semi-blind initialization} \label{sec:semi}

\subsubsection{Simulation Setup} \label{sec:setup}

The source signals were monochannel $16$~\!kHz signals randomly taken from the TIMIT database~\cite{garofolo93ti}. 
Each source signal was convolved with a binaural room impulse responses (BRIRs) from \cite{hummersone13a} to produce the corresponding ground truth source image. 
The images of the 3 or 4 sources were added to provide the mix signal. The BRIRs were recorded with a dummy head equipped with $2$ ear microphones, 
placed in a large lecture theatre of dimensions $23.5$~\!m $\times$ $18.8$~\!m $\times$ $4.6$~\!m, and reverberation time $\text{RT}_{60} \approx 0.68$~\!s \cite{hummersone13a}. 
We used a subset of (time-invariant) BRIRs with azimuthal source-to-head angle varying from $-90^\circ$ to $90^\circ$ with a $5^\circ$ step. Continuous circular movements 
were simulated by interpolating the BRIRs at the sample level using up-sampling, delay compensation, linear interpolation, delay restoration, and downsampling. 
Due to memory limitations, we truncated the original $16,\!000$-tap BRIRs to either $512$ or $4,\!096$ taps. Choosing two different lengths enables to evaluate the adequacy
 of the narrow-band assumption. 
Note that the recorded BRIRs almost vanish after $4,\!096$ samples, but not after $512$ samples. 

To assess the potential of the proposed algorithm to infer the time-varying frequency responses of the mixing filters, 
we devised two setups for the movement of the sources around the dummy head, drawn 
in Fig.~\ref{fig:trajectories}. In Type I mixtures, Source $s_3$  always goes from $85^\circ$ to $45^\circ$. The amplitude of the trajectory of all other sources is varied with $\vartheta \in \{15^\circ,30^\circ,45^\circ,60^\circ,75^\circ,90^\circ\}$. Each trajectory is covered at fixed speed, within the 
approximate $2$s of signal duration (all signals are truncated to $32,\!768$ samples).
We used four combinations of mixture type, filter tap length and number of sources, namely: \emph{I-512-3}, \emph{I-4096-3}, 
\emph{II-512-3},\footnote{In this case we discarded the fourth source (green plot in Fig.~\ref{fig:trajectories}).} and \emph{II-512-4}.
%

The STFT was applied to the mixed signal with a 512-sample, $50\%$-overlap, sine window, leading to $L = 128$ observation frames.
The number of components per source was set to $|\mathcal{K}_j| = 25$.
The correct number of sources in the mixture (3 or 4) was provided to the algorithms in all experiments, along with the component-to-source partition $\mathcal{K}$. The number of iterations for all methods was fixed to $100$.


\subsubsection{Performance measures}
Two standard audio source separation objective measures were calculated between the estimated and ground truth source images, 
namely: signal-to-distortion ratio (SDR) and signal-to-interference ratio (SIR) \cite{vincent06pe}.\footnote{We do not report and discuss signal-to-artefact ratio (SAR) measures in this subsection, due to space limitation.} In practice we used the \texttt{bss\_eval\_image} Matlab function dedicated to multichannel signals\footnote{http://bass-db.gforge.inria.fr/bss\_eval/.} \cite{vincent07fi}. 
Each reported measure is the average over $10$ experiments with different source signals, and different NMF initializations (see below).

\subsubsection{Baseline method}\label{sec:baseline}
The chosen baseline is a block-wise adaptation of the state-of-the-art method in \cite{ozerov10mu}. We adapted the implementation provided by the authors\footnote{http://www.unice.fr/cfevotte/publications.html.}, following the line described in the introduction. 
We first segmented the sequence of $L=128$ frames of the input mix into $P$ blocks of $L_p=L/P$ consecutive frames, 
and applied the baseline method to each block independently (i.e. to each $I \times F \times L_p$ subarray of mixture coefficients). 
Hence for each block we obtain a subarray of the source image STFT coefficients estimates.
Then by concatenating the successive subarrays and applying inverse STFT with overlap-add we obtain complete time-domain estimates of the 
source images.
As mentioned in the introduction, the block size $L_p$ must assume a good trade-off between local stationarity of mixing filters and a sufficient number of data 
to construct relevant statistics.
The method in \cite{ozerov10mu} was found to be very sensitive to the above constraint. For the simulations, we used $P=4$ ($\Leftrightarrow 
L_p=32$). This value showed better overall performance over the entire range of $\vartheta$.

\subsubsection{Initialization}\label{sec:init}
The proposed VEM requires initializing $\{w_{fk},h_{k\ell},\hat{\mathbf{a}}_{:,f\ell},$ $\symmat{\Sigma}^{\eta a}_{f\ell},\symmat{\Sigma}^a_f,\symvec{\mu}^a_f,\var_f\}_{f,\ell,k=1}^{F,L,K}$. The baseline method requires initializing $\{w_{fk},h_{k\ell},\mathbf{A}_f^p,\var_f\}_{f,\ell,k=1}^{F,L,K}$.
Note that all $P$ blocks share the same $w_{fk}$, each block has its own set of $\mathbf{A}_{f}^p, \var_f$ and also a subset of $h_{k\ell}$
(though an additional block index is omitted for clarity).

NMF parameters: The initial values for the NMF parameters $\{w_{fk},h_{k\ell}\}$, $k \in \mathcal{K}_{j}$ of a given source $j$ are calculated 
by applying the KL-NMF algorithm \cite{fevotte09no} to the monochannel power spectrogram of source $j$, with random initialization. In order to assess 
the robustness of the proposed method to ``realistic'' initialization, KL-NMF is applied to a corrupted version of the source spectrogram. For this, 
the time-domain source signal $s_j(t)$ is first summed with all other interfering source signals with a controlled signal-to-noise ratio (SNR) $R$. We 
tested three different levels of corruption, namely $R \in \{20\text{~dB},10\text{~dB},0\text{~dB}\}$, with $0$~dB meaning here equal power of signal 
$s_j(t)$ and of the sum of all interfering source signals. Note that $R=20$~dB is a quite favorable initialization, whereas $R=0$~dB tends towards 
more realism. This NMF initialization process is applied independently to all sources $j \in [1,J]$. The same resulting NMF initial parameters are 
used for both the proposed and baseline methods.

Mixing vectors: As for the initialization of $\hat{\mathbf{a}}_{:,f\ell}$, we used two different strategies. In the first one, for each source and 
each block of the baseline method, the time-interpolated BRIR corresponding to the center of the block was selected for the initialization of the 
corresponding column of $\mathbf{A}_f^p$ (after applying a 512-point FFT). For the proposed method, this initial $\mathbf{A}_f^p$ was replicated at 
each frame of the block, then vectorized, and set as initial $\hat{\mathbf{a}}_{:,f\ell}$. Applying this process to each block results in a complete 
initial sequence of $L$ mixing vectors $\hat{\mathbf{a}}_{:,f\ell}$. In the following, we refer to this strategy as \emph{Central-A}. The second 
strategy, called \emph{Ones-A}, consists of setting all the entries of $\mathbf{A}_f^p$ and $\hat{\mathbf{a}}_{:,f\ell}$ to $1$, $\forall f,\ell$. 
Obviously, this is a truly blind and challenging setup. Note that in all cases, both proposed and baseline algorithms were initialized with the same 
amount of filter information. 

Other parameters: The remaining parameters were initialized as follows: $\symmat{\Sigma}^{\eta a}_{f\ell} = 10^3 \mathbf{I}_{IJ}, \symvec{\mu}^a_f = 
\hat{\mathbf{a}}_{:,f1}, \symmat{\Sigma}^a_f = \mathbf{I}_{IJ}, \forall f,\ell$. As for the sensor noise variance $\var_f$, the baseline method 
showed the best performance when initialized with $1\%$ of the $(L,I)$-average PSD of the mixture, as suggested in \cite{ozerov10mu}. Our method 
behaved best with a much higher initial value for $\var_f$, namely $1,\!000$ times the $(L,I)$-average PSD of the mixture.

\subsubsection{Results} \label{sec:res}

\begin{figure}
\centerline{\includegraphics[width=0.95\columnwidth]{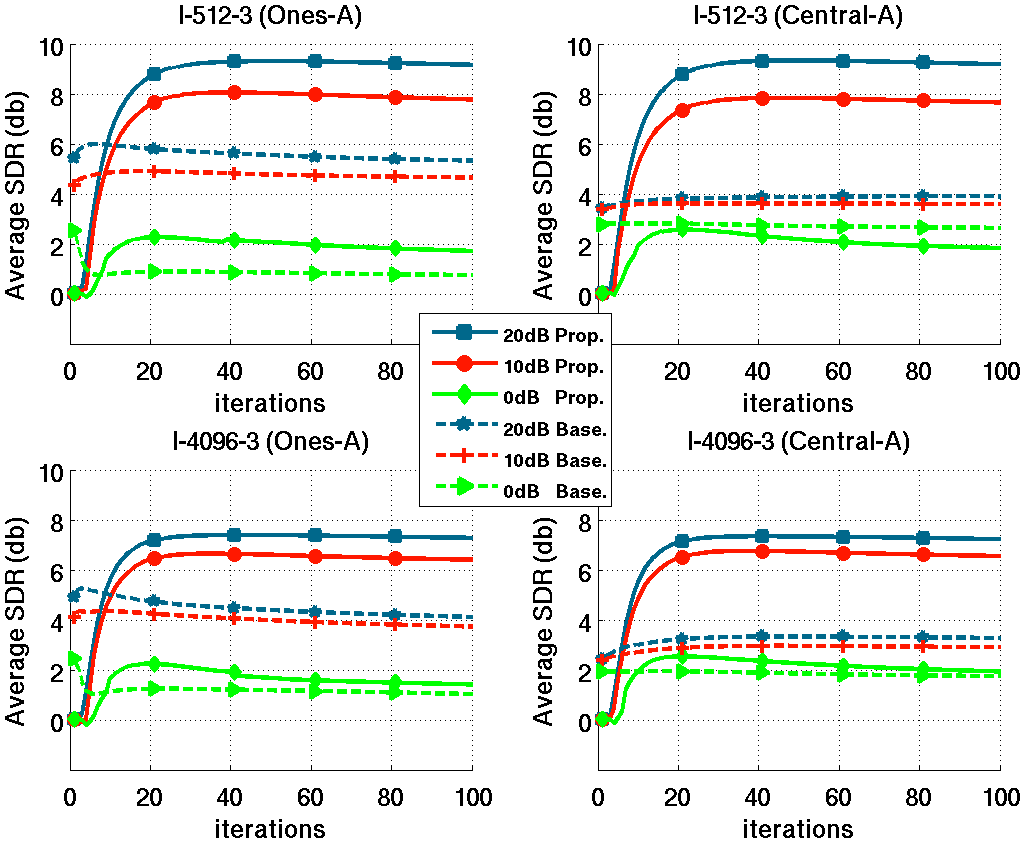}}
\caption{Overall-sources average SDR vs iterations. For different initialization schemes:
(top): \emph{I-512-3}, (bottom): \emph{I-4096-3}, (left) column is with \emph{Ones-A} initialization, (right) is with \emph{Central-A}. All 
experiments are at $\vartheta=75^\circ$.}
\label{fig:initialeogram}
\end{figure}

We first discuss detailed results for a particular (but representative) value of $\vartheta$, namely $\vartheta = 75^\circ$. Then we
report the performance of the proposed ASS algorithm w.r.t. the variation of $\vartheta$ and generalize the discussion.  

Fig.~\ref{fig:initialeogram} represents the evolution of average SDR measures with the (V)EM iterations, for $\vartheta = 75^\circ$, and Mix-I. Let us 
recall that SDR is a general indicator that balances separation performance (i.e. interfering source rejection) and signal distortion (reconstruction 
artifacts). Each line is the result of averaging over the 3 sources, and over 10 different runs with different source signals. The two upper plots 
correspond to mix \emph{I-512-3} and the two lower plots correspond to mix \emph{I-4096-3}. The two left plots were initialized with the \emph{Ones-A} 
strategy and the two right plots were initialized with \emph{Central-A}. 

In a general manner, the curves show that the baseline method converges faster than the proposed method, which is natural since the baseline method 
functions on blocks of STFT frames and the proposed method uses the complete sequence of STFT frames. Also, the baseline method has less parameters to 
estimate. In \emph{I-512-3} (\emph{Central-A}), the proposed method has an average performance of SDR $\approx 9.5$~dB for $R=20$~dB. The SDR score 
slightly degrades to about $8$~dB for $R=10$~dB, and then more abruptly decreases to about $2$~dB for $R=0$~dB. SDR scores of the baseline method at 
$R=20$~dB, $10$~dB, and $0$~dB go from $4$ to $2.5$~dB. Therefore, the proposed VEM largely outperforms the baseline method for $R=20$~dB and $10$~dB, 
though in this example, the baseline performs slightly better at $R=0$~dB ($\approx +0.5$~dB over the proposed method). 

Regarding the influence of the initialization of the mixing vectors initialization, \emph{Ones-A} vs. \emph{Central-A}, the proposed algorithm 
proves to be remarkably robust to poor mixing filter initialization, since \emph{Ones-A} provides similar results to \emph{Central-A}. 
Hence, the 
proposed algorithm is able to correctly infer the mixing vectors from blind initialization, given that some reasonable amount of information on 
source PSD is provided (for instance by the NMF initialization). As for the baseline, its scores for $R=20$~dB and $10$~dB are again largely below the 
scores of the proposed method. 
However, and quite surprisingly, the baseline method behaves better (by about $0.4$--$0.7$~dB) in the \emph{Ones-A} (blind) configuration compared to 
the \emph{Central-A} configuration, for $R=20$~dB and $10$~dB. This result is a bit difficult to interpret, but a possible explanation is that we 
measure the performance using the source images, rather than the monochannel source signals. Nevertheless for $R=0$~dB, the filter information 
delivered by \emph{Central-A} seems more useful, since the performance of the baseline method in the \emph{Ones-A} configuration is about $2$~dB 
lower than for \emph{Central-A}. As a result, in the \emph{Ones-A} configuration, the SDR scores of the proposed VEM are above the scores of the 
baseline method for all tested $R$ values, including $R=0$~dB.

As for the influence of the length of the BRIRs, we see that, unsurprisingly, the performance of both proposed and baseline algorithms 
decreases when the BRIRs go from $512$-tap to $4096$-tap responses.
For $R=20$~dB and $10$~dB, we can observe that the decrease is of about $1.5$--$2$~dB for the proposed method, independently of the mixing vectors 
initialization. 
The decrease is lower for the baseline method ($\approx 1$~dB), but this is probably related to the fact that the baseline scores are lower. 
For $R=0$~dB, the influence of the BRIRs length on the performance of the proposed method is quite moderate, but this is also probably because the SDR 
scores are much lower than for $R=20$~dB and $10$~dB. 
All this manifests that \eqref{eq:temp1} becomes a less appropriate model as the reverberation increases. 
Note that this is a recurrent problem in ASS in general. 
Our VEM is not intended to deal with this problem, but these experiments show that our VEM can provide quite remarkable SDR scores in a configuration 
that is \emph{very} difficult in many aspects (underdetermined, time-varying, reverberant).

\begin{table*}[!t]
\centering
\caption{Average SDR and SIR measures for $\vartheta=75^\circ$, \emph{Ones-A}.}
\label{tab:quantitative}
\resizebox{0.99 \linewidth}{!}{%
\begin{tabular}{c c c cccc c cccc c cccc  c cccc}
\toprule
 &&& \multicolumn{9}{c}{SDR} && \multicolumn{9}{c}{SIR} \\
 \midrule
 &&& \multicolumn{4}{c}{Proposed} && \multicolumn{4}{c}{Baseline} && \multicolumn{4}{c}{Proposed} && 
\multicolumn{4}{c}{Baseline}\\[1.1ex]
$R$ & Mixture && $s_1$ & $s_2$ &  $s_3$ & $s_4$  && $s_1$ & $s_2$ &  $s_3$ & $s_4$ && $s_1$ & $s_2$ & $s_3$ & $s_4$ && $s_1$ & $s_2$ & $s_3$ & $s_4$\\
\midrule
\multirow{4}{*}{$20$~\!dB} &
I-512-3  &&      \textbf{9.3} & \textbf{10.4} & \textbf{7.9}  & -- &&          5.5  & 6.5 & 4.0  & -- &&
		 \textbf{14.9} & \textbf{16.0} & \textbf{14.3} & --  &&          10.5 & 12.3 & 8.4  & --  \\ 
& I-4096-3 &&      \textbf{7.7} & \textbf{7.9}  & \textbf{6.2}  & -- &&          4.7  & 4.6 & 3.0  & -- &&  
		 \textbf{13.0} & \textbf{13.7} & \textbf{11.3} & --  &&          10.0 & 9.9  & 6.6  & --  \\ 
& II-512-3 &&      \textbf{8.4} & \textbf{8.2}  & \textbf{9.5}  & -- &&          4.4  & 4.5 & 5.7  & -- &&  
		 \textbf{13.6} & \textbf{13.8} & \textbf{16.1} & --  &&          8.6  & 9.1  & 12.2 & --  \\ 
& II-512-4 &&      \textbf{7.0} & \textbf{6.6}  & \textbf{7.6}  & \textbf{9.2} &&  3.8  & 3.9 & 4.9  & 5.8 && 
		 \textbf{11.4} & \textbf{11.8} & \textbf{14.2} & \textbf{15.7} &&  7.4  & 8.7  & 9.8  & 11.3 \\ 
\midrule
\multirow{4}{*}{$10$~\!dB} &
I-512-3  &&      \textbf{7.9} & \textbf{9.1}  & \textbf{6.3}  & -- &&          4.8  & 6.0 & 3.1  & -- && 
		 \textbf{12.8} & \textbf{13.6} & \textbf{12.9} & --  &&          9.4  & 11.5 & 7.2  & -- \\
& I-4096-3 &&      \textbf{6.9} & \textbf{7.1}  & \textbf{5.2}  & -- &&          4.2  & 4.4 & 2.5  & -- && 
		 \textbf{11.4} & \textbf{11.7} & \textbf{9.7}  & --  &&          9.0  & 9.2  & 5.7  & -- \\ 
& II-512-3 &&      \textbf{7.1} & \textbf{6.9}  & \textbf{8.2}  & -- &&          3.8  & 4.0 & 5.3  & -- && 
		 \textbf{11.5} & \textbf{12.2} & \textbf{13.9} & --  &&          7.5  & 8.5  & 11.3 & -- \\ 
& II-512-4 &&      \textbf{6.1} & \textbf{6.0}  & \textbf{6.9}  & \textbf{8.2} &&  3.7  & 3.9 & 4.6  & 5.4 &&
		 \textbf{10.4} & \textbf{10.6} & \textbf{12.8} & \textbf{13.7} && 6.8  & 8.1  & 8.8  & 10.7 \\
\midrule
\multirow{4}{*}{$0$~\!dB} &
I-512-3  &&      \textbf{2.4} & \textbf{2.7}  & \textbf{0.0} & -- &&          1.1  & 2.3 & -1.2 & -- &&
		 \textbf{4.3} & 4.4 & -0.4 & -- &&             3.7  & \textbf{5.9}  & \textbf{0.0}  & -- \\ 
& I-4096-3 &&      \textbf{2.0} & 1.9  & \textbf{0.3}  & -- &&          1.8  & \textbf{2.1} & -0.8 & -- && 
		 4.2 & 3.6 & \textbf{-0.2} & -- &&             \textbf{4.9}  & \textbf{5.1}  & -0.5 & -- \\ 
& II-512-3 &&      \textbf{1.1} & \textbf{1.1}  & \textbf{2.7}  & -- &&          0.0 & 0.4 & 1.7  & -- && 
		 \textbf{2.5} & 2.1 & 3.9  & -- &&             2.0  & \textbf{3.3}  & \textbf{4.2}  & -- \\ 
& II-512-4 &&      \textbf{1.8} & \textbf{1.7}  & \textbf{3.4}  & \textbf{3.8} &&  0.7  & 1.0 & 1.7  & 2.3 &&
		 \textbf{4.2} & \textbf{3.6} & \textbf{5.3}  & \textbf{5.8} &&  2.7  & 3.2  & 3.3  & 4.6 \\ 
\bottomrule
\end{tabular}}
\end{table*}

\begin{table}[h!]
\caption{Input SDR and SIR for the $4$ different mixtures.}
\label{tab:input}
\centering
\resizebox{\columnwidth}{!}{%
\begin{tabular}{c cccc cccc}
\toprule
 & \multicolumn{4}{c}{SDR} & \multicolumn{4}{c}{SIR}\\
Mixture & $s_1$ & $s_2$ &  $s_3$ & $s_4$  & $s_1$ & $s_2$ &  $s_3$ & $s_4$ \\
\midrule
I-512-3  &  -3.4 & -1.2 & -7.6 & -- & -2.0 & -0.5 & -5.9 & --  \\
I-4096-3 &  -2.6 & -2.0 & -7.5 & -- & -2.0 & -0.5 & -5.9 & --  \\
II-512-3 &  -5.3 & -4.9 & -2.1 & -- & -4.1 & -3.7 & -1.1 & --  \\ 
II-512-4 &  -7.8 & -7.6 & -5.3 & -4.1 & -6.3 & -6.0 & -4.1 & -3.5  \\ 
\bottomrule
\end{tabular}}
\end{table}


Table~\ref{tab:quantitative} provides results (at iteration $100$) that are detailed per source (still averaged over 10 mixtures), and extended to 
SIR, for $\vartheta=75^\circ$ and \emph{Ones-A} filter initialization. Output SIR scores focus on the ability of an ASS method to reject interfering 
sources. We first see from Table~\ref{tab:quantitative} that for $R=20$~dB and $R=10$~dB, the proposed VEM outperforms the baseline in both SDR and 
SIR 
for all configurations. In other words, the hierarchy discussed when analyzing Fig.~\ref{fig:initialeogram} for $R=20$~dB and $R=10$~dB extends to 
per-source results, to Mix-II, and to SIR (at least for \emph{Ones-A}). SDR improvement of the proposed method over the baseline ranges from $2.1$~dB 
($s_2$ in \emph{II-512-4} at $R=10$~dB) to $4.0$~dB ($s_1$ in \emph{II-512-3} at $R=20$~dB). SIR improvement of the proposed method over the baseline 
ranges from $2.1$~dB ($s_2$ in \emph{I-512-3} at $R=10$~dB) to an impressive $5.9$~dB ($s_3$ in \emph{I-512-3} at $R=20$~dB). The results are 
particularly 
remarkable for the 4-source mixture configuration, with a range of output score similar to the 3-source configuration, and improvement over the 
baseline method up to $4.4$~dB ($s_3$ and $s_4$ at $R=20$~dB).
At $R=0$~dB the SIR results are more deteriorated for the 3-source configurations: they do not seem to indicate which method performs best (in terms 
of SIR). 
However, the SDR scores at $0$~dB are all higher for the proposed method than for the baseline method, except for $s_2$ in mixture \emph{I-4096-3} 
(only $0.2$~dB below the baseline though). The improvement is however more limited than for $R=20$~dB and $R=10$~dB (maximum improvement is here 
$1.3$~dB). Finally, at $R=0$~dB, it can be noted that for the 4-source mixture, the proposed method outperforms the baseline method for all sources, 
and for both SDR (improvement ranges from $0.7$~dB to  $1.7$~dB) and SIR (improvement ranges from $0.4$~dB to $2$~dB).

For a given source, the performance of ASS is more adequately described by the \emph{separation gain}, i.e. the difference between output score and 
input score than by the output score only. Indeed, an input score quantifies how much the target source is corrupted in the input mixture. A source 
with low input score is more difficult to extract than a source with high input score. We thus display in Table~\ref{tab:input} the input SDR and 
input SIR scores of each source.\footnote{We can see in this table that the length of BRIRs does not affect the input SIR, i.e. the entries 
\emph{I-512-3} and \emph{I-4096-3} are the same up to $2^\text{nd}$ decimal figure), when it slightly degrades the corresponding SDR scores.} 
Subtracting the scores in Table~\ref{tab:quantitative} and Table~\ref{tab:input}, we can obtain SDR gains and SIR gains. We comment the results for 
$R=0$~dB since it is the most realistic setting (remind that we also are in the \emph{Ones-A} blind configuration for filters). For the 3-source 
mixtures, the proposed VEM algorithm provides a SDR gain ranging from $3.9$~dB to $7.8$~dB, and an SIR gain ranging from $4.1$~dB to $5.8$~dB.
As for the 4-source mixture, it is interesting to see that sources $s_3$ and $s_4$ score higher than $s_1$ and $s_2$ in Table~\ref{tab:quantitative}, 
although they move twice as fast as $s_1$ and $s_2$ and are thus expected to be more difficult to separate. However, they also have higher input 
scores, so that the separation gain turns out to be quite similar across sources.

\begin{figure}
\centerline{\includegraphics[width=0.95\columnwidth]{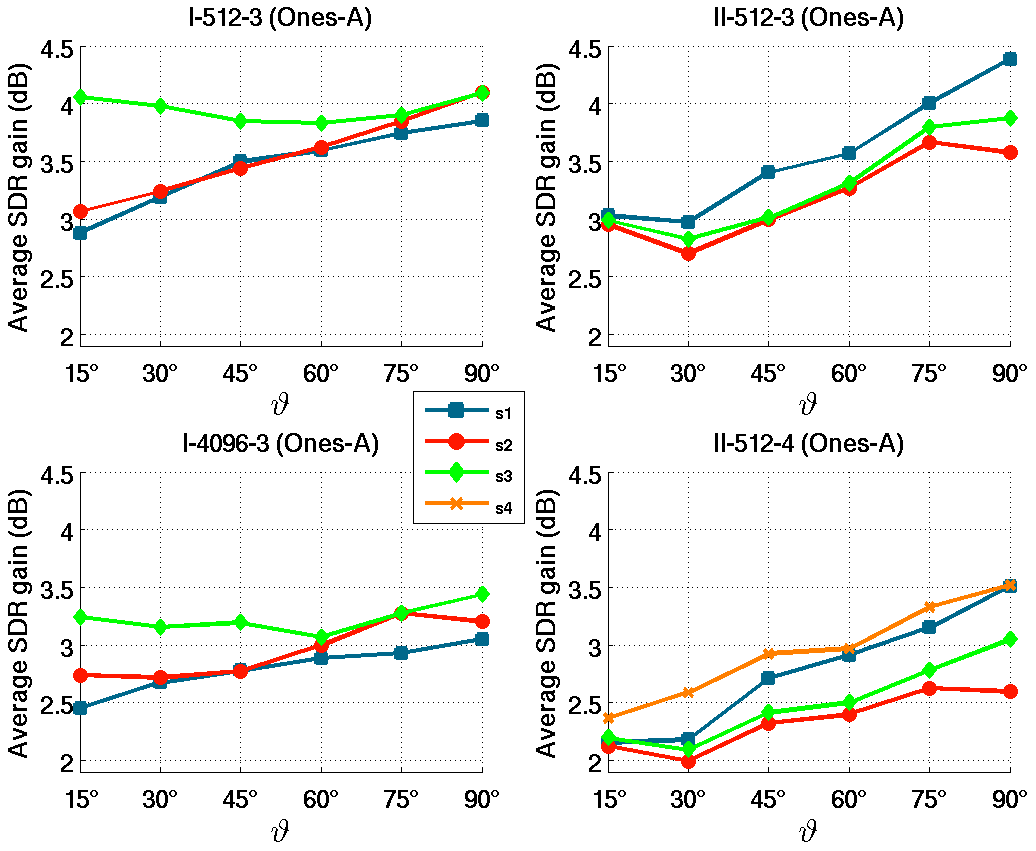}}
\caption{Average SDR gain of the proposed method over the baseline method, for the $4$-source mixture, as a function of  $\vartheta$ ($R=20$~dB, 
\emph{Ones-A} initialization).}
\label{fig:angleogram}
\end{figure}

We now focus on performance behavior w.r.t. the source velocity, i.e different values of $\vartheta$. Fig.~\ref{fig:angleogram} plots the gain of the 
proposed method over the baseline method, i.e. the (signed) difference of the proposed method's SDR and the SDR of the baseline. The results shown in 
Fig. \ref{fig:angleogram} are at $R=20$~\!dB, and \emph{Ones-A} strategy (as the latter was shown to be most favorable for the baseline).
For \emph{II-512-3}, we observe that except for the 3 sources at $\vartheta = 30^\circ$ and for $s_2$ at $\vartheta = 90^\circ$, the 
gain is monotonically increasing for all three sources, starting from about $3$~dB at $\vartheta = 15^\circ$ and going up to $3.5$--$4.5$~dB at 
$\vartheta = 90^\circ$. Therefore, the advantage of the proposed method over the block-wise approach gets larger as the speed of moving sources 
increases. This makes sense since the block-wise baseline method rely on the assumption that filters are stationary on each block, and this assumption 
gets mangled as the source speed increases. In contrast, the proposed method seems robust to a large range of source velocity.
This trend is also visible on the other plots. For example, for the \emph{I-512-3} mixture, we see that the gain increases with $\vartheta$ for $s_1$ 
and $s_2$, from about $3$~dB at $\vartheta = 15^\circ$ to about $4$~dB at $\vartheta = 90^\circ$, whereas the gain for $s_3$ (whose trajectory remains 
independent of $\vartheta$) is almost constant at about $4$~dB. The decreasing of this latter curve a bit around $\vartheta = 45^\circ$ may be due to 
the trajectories of $s_1$ and $s_2$ interfering with the trajectory of $s_3$ for $\vartheta \ge 45^\circ$. Additionally, the $s_3$ curve in 
configuration \emph{I-512-3} shows that the advantage of the proposed method can be also large for relatively slow sources.

\begin{table*}[t!]
\caption{Average measures using blind initialization, for simulations and real recordings (all units are dB).}
\label{tab:blindinit}
\resizebox{0.99 \linewidth}{!}{%
\begin{tabular}{c c c ccc c ccc c ccc  c ccc     c ccc}
\toprule
 &&& \multicolumn{7}{c}{simulated Mix-270} && \multicolumn{7}{c}{simulated Mix-680} &&  \multicolumn{3}{c}{real recordings} \\
 \midrule
SNR    &     && \multicolumn{3}{c}{$\infty$} && \multicolumn{3}{c}{$4$}     &&      \multicolumn{3}{c}{$\infty$} && \multicolumn{3}{c}{$4$}        && 
\multicolumn{3}{c}{N/A}\\[1.1ex]
Method & Src &&            SDR &  SIR & SAR  &&       SDR &  SIR & SAR      &&        SDR &  SIR & SAR           && SDR &  SIR & SAR               && 
SDR & SIR & SAR \\
\midrule
\multirow{3}{*}{Input} 
& $ s_1 $   &&   -2.3  & -1.9 & $+\infty$   &&    -4.5  & -1.9 & 4.6   &&     -3.5  & -2.9 & $+\infty$   &&    -5.5  & -2.9 & 4.6  
&& 0.0   &  0.2  &  $+\infty$ \\ 
 & $ s_2 $   &&   -3.8  & -3.0 & $+\infty$   &&    -5.7  & -3.0 & 4.6   &&     -2.7  & -1.9 & $+\infty$   &&    -4.8  & -2.0 & 4.6 
&&  0.0    &  0.2  &  $+\infty$ \\ 
 & $ s_3 $   &&   -3.1  & -2.5 & $+\infty$   &&    -5.1  & -2.6 & 4.6   &&     -3.3  & -2.7 & $+\infty$   &&    -5.3  & -2.7 & 4.6 
&&  -   &  -  &  - \\
\midrule
\multirow{3}{*}{Bin-Mask} 
 & $ s_1 $   &&   6.2  & 10.5 & 9.5   &&    2.5  & 7.5 & 3.4   &&     2.8  & 5.2 & 6.1   &&    0.5 &  2.6 & 1.7  
&&  2.9   &  7.6  &  6.3\\ 
 & $ s_2 $   &&   6.2  & 10.8 & 9.4   &&    2.0  & 6.9 & 3.4   &&     3.8  & 6.9 & 8.2   &&    1.2 &  4.7 & 3.1
 &&  3.1    &  6.4  &  6.6\\ 
 & $ s_3 $   &&   5.9  & 9.9 & 9.2   &&    1.9  & 6.0 & 3.0   &&     2.6  & 3.8 & 6.8   &&    0.7 &  2.7 & 2.7
 && -    &  -  &  -\\
\midrule
\multirow{3}{*}{Baseline}
 & $ s_1 $   &&    6.0      & 11.1 & 9.7             &&                3.2  & 7.9 & 5.3   &&     2.3  & 4.9 & 6.4   &&    0.7  & 2.6 & 3.4          
 && 3.5    &  6.7  &  \textbf{8.3}\\ 
& $ s_2 $   &&   6.7       & 11.1 & 10.0         &&               2.9  & 7.7 & 5.0                &&     3.8  & 7.1 & 8.5   &&    1.6  & 4.9 & 4.4 
&& 3.6    &  6.1  &  9.1\\ 
& $ s_3 $   &&   5.9         & 9.7 & 9.5       &&               2.8  & 6.7 & 4.8                    &&     2.5  & 4.4 & 7.1   &&    1.1 & 2.8 & 4.2  
&& - & - & -\\
\midrule
\multirow{3}{*}{Proposed} 
 & $ s_1 $   &&   \textbf{7.5}  & \textbf{13.4}  & \textbf{11.5}   &&    \textbf{5.0}     & \textbf{10.0} & \textbf{8.9}       &&     \textbf{3.3}  & 
\textbf{6.8} & \textbf{7.8}   &&   \textbf{1.9} & \textbf{4.0} & \textbf{6.3}
&& \textbf{4.2}    &  \textbf{7.8}  &  \textbf{8.3} \\ 
 & $ s_2 $   &&   \textbf{7.8}  & \textbf{13.4}  & \textbf{11.7}   &&    \textbf{4.4}   & \textbf{9.4} & \textbf{8.5}        &&     \textbf{4.4}  & 
\textbf{8.3} & \textbf{9.6}   &&    \textbf{2.6}  & \textbf{5.7} & \textbf{7.4}  
&& \textbf{4.5}    &  \textbf{7.1}  &  \textbf{9.2}\\ 
 & $ s_3 $   &&   \textbf{7.4}  & \textbf{11.7} &  \textbf{11.3}   &&    \textbf{4.6}   & \textbf{7.9} & \textbf{8.5}      &&     \textbf{3.0}  & 
\textbf{4.9} & \textbf{8.2}   &&    \textbf{2.3}  & \textbf{3.4} & \textbf{7.3}  
&&  -    &  -  &  -\\  
\bottomrule
\end{tabular}}
\end{table*}

\subsection{Experiments with blind initialization} \label{sec:blind}

\addnote{1}{In this section, we report the second series of experiments, that were conducted with blind initialization. This series of experiments 
consists of two parts: the first part deals with simulated 3-speaker mixtures, and the second part deals with a 2-speaker mixture made of 
real recordings. We first present the blind initialization method, that is common to all these new experiments, and then we detail the set-ups and 
results in the next subsections.}

\subsubsection{Blind initialization}\label{sec:init2}

\addnote{1}{In these new experiments, the initialization of the proposed VEM algorithm (and of the baseline method) relies on the use 
of a state-of-the art blind source separation method based on source localization and binary masking. More specifically, we adapted the sound source 
localization method of \cite{dorfan15tr}, which is a good representative of recently proposed probabilistic methods based on mixture models of 
acoustic feature distribution parameterized by source position, see e.g. \cite{mandel10mo, may11ap, woodruff12bi, traa14mu}. The method in
\cite{dorfan15tr} relies on a mixture of complex Gaussian distributions 
(CGMM) that is used to compare the measured normalized relative transfer function (NRTF) at a pair of microphones with the expected NRTF 
as predicted by a source at a candidate position and a direct-path propagation model (there is one CGMM component for each candidate source position 
on a predefined grid). Combining the measures obtained at different microphone pairs into an EM algorithm enables 
to estimate the priors of the CGMM components. Then selecting the $J$ first maxima of the priors amounts to localize the $J$ sources. 
It also delivers the associated mixing vectors (corresponding to the direct path between sources and microphones). We adapted this method to the use 
of one pair of microphone, delivering $J$ 
source direction estimates (in azimuth) and corresponding mixing vectors. We further combined it with a binary mask for source separation, inspired by 
\cite{dorfan15sp}. For each TF bin, the masks are obtained by comparing the 
measured NRTF with the NRTF corresponding to the $J$ candidate source directions; the source obtaining the largest posterior value in the CGMM among 
the $J$ selected components has its mask set to 1 while the other sources have their mask set to 0. Then for each source, the mask is classically 
applied to the mixture STFT to obtain an estimate of the corresponding source image STFT. Importantly, to deal with our time-varying mixing 
set-up, this process is applied in a block-wise mode, similarly to what is done with the baseline method (see Section~\ref{sec:baseline}). Mixing vectors estimated on each block are replicated 
and catenated to form the initial $\hat{\mathbf{a}}_{:,f\ell}$ $L$-sequence. For each source $j$, the block-wise estimates of source image STFT 
vectors obtained by the binary masking are also concatenated, transformed to absolute squared values, averaged across channels, and supplied to the 
KL-NMF algorithm \cite{fevotte09no} to provide initial NMF parameter estimates for the complete sequence of $L$ frames.
This blind source separation method has been shown to be robust to short blocks, and therefore we can use here more blocks (of course shorter blocks) 
than in Section~\ref{sec:baseline}. This method was thus applied with $16$ blocks (to process 2-second signals, with $50\%$ overlap, hence one block 
is $250$~ms long). Note that the baseline method that is plugged onto the initialization method is still run with $P=4$ blocks. Note also that, as in 
Section~\ref{sec:semi}, the same information is used for the initialization of the proposed VEM and for the initialization of the baseline method.}

\subsubsection{Simulation set-up}\label{sec:setup2}

\begin{figure}[t]
\centering
\centerline{\includegraphics[width=0.45\columnwidth]{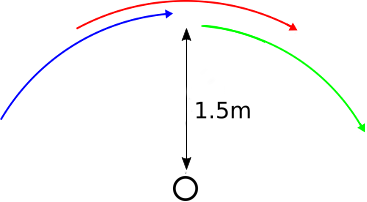} \hspace{1cm}
\includegraphics[width=0.45\columnwidth]{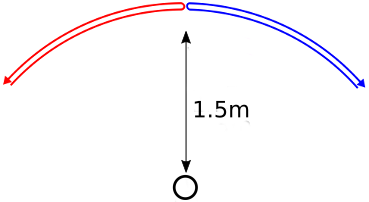}}
\caption{Source trajectories for the experiments with blind initialization: Simulations (left) and real recordings (right).}
\label{fig:trajectories2}
\end{figure}

\addnote[]{1}{The new simulation set-up is an underdetermined stereo setup of $J=3$ simulated moving speakers (two male and one female from TIMIT).
Since the blind initialization method relies on a free-field direct-path propagation model, we replaced the dummy head binaural recordings of Section~\ref{sec:semi} with the room impulse response (RIR) simulator of AudioLabs Erlangen,\footnote{available at 
www.audiolabs-erlangen.de/fau/professor/habets/software/rir-generator.} based on the image method \cite{allen79im}. We defined a 2-microphone set-up 
with omnidirectional microphones, spaced by $d=50$~cm. The simulated room had the same size as the one in Section~\ref{sec:setup}.
In Section~\ref{sec:setup}, we had simulated sources trajectoires that were crossing multiple times, to test the proposed method in a difficult scenario.
However, the binary-mask initialization method is applied on blocks of time-frames, and it may be subject to source permutation across blocks.\footnote{Note however that it is not subject to source permutation across frequency bins since all frequencies are jointly considered in the CGMM model, see \cite{dorfan15tr} for details.} To avoid this problem, we simulated a new setup where the trajectories of the $J=3$ sources are not crossing each other: The 3 speech sources are all moving in circle of $\vartheta = 60^\circ$ in $2$~s, from $-65^\circ$ to $-5^\circ$ for $s_1$, from $-30^\circ$ to 
$30^\circ$ for $s_2$ and from $5^\circ$ to $65^\circ$ for $s_3$, at about $1.5$~m of the microphone pair center (see Fig.~\ref{fig:trajectories2}-left).
We simulated two reverberation times, namely $T_{60}=680$~ms (same as in Section~\ref{sec:semi}) and $T_{60}=270$~ms (the corresponding mixtures are 
denoted respectively as \emph{Mix-680} and \emph{Mix-270}). We also tested the mixtures as is (noiseless case) and 
corrupted with additive white Gaussian noise (AWGN) at SNR$=4$~dB. This resulted in $4$ configurations. All reported measures are average results 
over 10 mixtures using different speech signals from TIMIT.}

\subsubsection{Real recordings set-up}\label{sec:setup3}

\addnote{1}{Real recordings were made in a $20$ m$^2$ reverberant room ($T_{60} \approx 500$~ms), using $I=2$ omnidirectional microphones in 
free field, placed in the center of the room, and spaced by $d=30$~cm. For real recordings, the blind initialization method was shown to be much less 
efficient to separate 3 speakers, compared to the simulated experiments, but still worked very well for 2 speakers. We thus limited the present 
experiments with 2 speakers. Two speakers (one female, one male) were thus asked to pronounce spontaneous speech while moving on a circle at $1.5$~m 
from the microphones, of about $45^\circ$, two-way opposite motions, starting respectively at about $45^\circ$ and $-45^\circ$ (see Fig.~\ref{fig:trajectories2}-right). The trajectory was traveled within 2 s, hence the speaker movement was pretty fast. The two speakers were recorded separately, and the signals were added, so that we could calculate separation scores.}

\subsubsection{Results of simulations}\label{sec:res2}
\addnote{1}{Measures are reported in Table~\ref{tab:blindinit} for the input mixed signals, the initial source estimates after the binary masking, the 
estimates using the baseline method and the estimates using the proposed method. In addition to the SDR and SIR measures, we also report here 
signal-to-artifacts ratios (SAR) which measure the quantity of artefacts introduced on the separated signal by the separation method. Note that relatively homogeneous input SDR scores across sources (around $-3$~dB and $-5$~dB for the noiseless and noisy case 
respectively for both \emph{Mix-270} and \emph{Mix-680}) indicate that all sources have roughly the same power in the mix.}

\addnote{1}{Let us start with the most reverberant condition  \emph{Mix-680}. At SNR~$=\infty$, the average SDR (across sources) attained by the 
binary masking 
method  is approximatively $3$~dB, hence a SDR gain of about $6$~dB over input signals. The corresponding average SIR gain is $7.8$~dB, and the output 
average SAR is  about $7$~dB.\footnote{It make poor sense to provide SAR gain, since, as source signals are intact in the mix, the input SAR is 
$=\infty$ and source separation can only lead to SAR decrease.} For this setting, the baseline method does not seem able to efficiently exploit the 
information provided by the blind initialization: The overall performance is comparable to the binary masking (SDR is even very slightly decreased 
for two sources). Regarding the proposed method, there is a significant improvement over both the binary mask initialization and the baseline method. 
In detail, the proposed method outperforms the baseline method by $0.5$~dB to $1$~dB SDR, by $0.5$~dB to $1.9$~dB SIR, and by $1.1$~dB to $1.4$~dB SAR 
(averaged across sources). With the addition of noise (SNR~$=4$~dB), all performance measures drop significantly, which was expected. For example, the 
average SDR for the binary masking is $2.3$~dB lower than for the noiseless condition. Here, the baseline method slightly improves the binary masking 
scores, by $0.3$~dB SDR, $0.1$~dB SIR, and $1.5$~dB SAR. More importantly, the proposed method outperforms the baseline method by $1.1$~dB SDR, 
$0.9$~dB SIR, and $3$~dB SAR. Note that under noisy conditions, there is more margin for improvement over the binary masking since the latter provides 
worse estimates than in the noiseless case. }

\addnote{1}{For \emph{Mix-270}, i.e. moderate reverberations, we obtain significantly higher separation scores for all methods, as expected. For 
example, at 
SNR~$=\infty$, the SDR for the binary masking (averaged across sources) is about $6$~dB, hence a SDR gain of about $9$~dB over input signals. Output SIR and SAR are 
within $9.2$~dB to $10.8$~dB (with a SIR gain going up to $13.8$~dB). These scores (the SIR measures in particular) confirm what is well-known in the 
literature: Binary-masking techniques show good separation performance in low-to-moderate reverberant conditions. They place our block-wise binary masking method at the level of state-of-the-art methods based on the same principles (two-microphone source localization and binary masking), e.g. \cite{mandel10mo, 
may11ap, woodruff12bi, traa14mu}, even though it is applied on quite short blocks ($250$~ms of mixture signal). Again, the baseline method exhibits comparable scores with the binary masking, here slightly better on the average. 
In addition, the proposed method significantly outperforms the baseline method, by $1.4$~dB SDR, $2.2$~dB SIR, and $1.8$~dB SAR. The proposed method 
obtains SIR gains with respect to inputs as high as $16.4$~dB (source $s_2$), which, we believe, is remarkable in a blind, underdetermined, dynamic 
setup, be it simulated. At SNR $=4$~dB, we observe the same trends as for \emph{Mix-680}: the baseline method improves more neatly over the binary 
masking, and the proposed method, again, significantly improves over the baseline method (by $1.7$~dB SDR, $1.7$~dB SIR, and $3.6$~dB SAR).}

\subsubsection{Results of real recordings}\label{sec:res3}
\addnote{1}{The last three columns of Table~\ref{tab:blindinit} report the performance measures obtained on the real recordings with two sources. We 
first notice 
that even if we mix two sources instead of three, the gain performance of the binary masking method is less notable that in our simulated scenarios. 
This is evidence that separating (two) moving sources from real recordings remains quite a challenging scenario, even for state-of-the-art sound 
processing techniques. The baseline method shows some SDR improvement ($\approx 0.5$~dB) and SAR improvement ($> 2$~dB) for both sources over the 
binary masking. However, the baseline SIR scores degrade when compared to the binary-masking initialization. The proposed method exhibits positive 
gains when compared both with the binary-masking initialization and with the baseline method. Indeed, SAR scores of the proposed method are 
equivalent to 
the baseline method and notably better than the initialization. SDR improves by more than $1$~dB when compared to the initialization, and by $0.7$~dB 
to $0.9$~dB when compared to the baseline method. SIR improves by $0.2$~dB to $0.7$~dB when compared to the initialization and by $0.7$~dB to 
$1.1$~dB when compared to the baseline method. Such results demonstrate the potential of the proposed approach for real-world 
applications and encourage us to pursue this line of research.}

\section{Conclusion and Future Work}
\label{sec:conclusion}

In this paper we addressed the challenging task of separating audio sources from underdetermined time-varying convolutive mixtures. We started with 
the multichannel time-invariant convolutive LGM-NMF framework of \cite{ozerov10mu}, and we introduced time-varying filters modeled by a first-order 
Markov model with complex Gaussian observation and transition distributions. Because the mixture observations do not depend only on the filters, but 
also on the sources that are latent variables as well, a standard direct application of a Kalman smoother is not possible. We addressed this issue 
with a variational approximation, assuming that the filters and the sources are conditionally independent with respect to the mixture. This lead to a 
closed-form variational EM (VEM), including a variational version of the Kalman smoother, and finally, separating Wiener filters that are constructed 
from both time-varying estimated source parameters and time-varying estimated mixing filters. Several implementation issues were discussed to 
facilitate experimental reproducibility. Finally, an extensive evaluation campaign demonstrated the experimental advantage of the proposed approach 
over a state-of-the-art baseline method in several speech mixtures under different initialization strategies.

These results encourage for further research to improve the proposed model. 
Firstly, the last series of reported experiments show that the use of realistic blind separation methods for the initialization of our algorithm in the case of more sources than microphones has to be more deeply explored and made more robust to process real recordings.
Secondly, in the present study, the number of sources present in the mixture was assumed to be known, although the estimation of this number is a problem on its own. Therefore, developing algorithms capable of estimating the number of active (i.e. emitting) sources varying over time remains an open issue, but is a step closer to realistic applications. We therefore plan to incorporate into the present model the estimation of the sources activity, using diarization latent 
variables. 
Finally, an in-depth study exploring the complex relationship between the physical changes of the recording set-up and the mixing filters 
can be of great help. In particular, a better understanding of how the position of the sources and microphones affect the 
filters may enable us to incorporate the rationale of the discrete DOA-dependent model in~\cite{higushi12un} to the proposed continuous latent 
model, thus using localization cues to help the automatic separation of sound sources.

\bibliographystyle{IEEEtran}

\begin{thebibliography}{10}
\providecommand{\url}[1]{#1}
\csname url@samestyle\endcsname
\providecommand{\newblock}{\relax}
\providecommand{\bibinfo}[2]{#2}
\providecommand{\BIBentrySTDinterwordspacing}{\spaceskip=0pt\relax}
\providecommand{\BIBentryALTinterwordstretchfactor}{4}
\providecommand{\BIBentryALTinterwordspacing}{\spaceskip=\fontdimen2\font plus
\BIBentryALTinterwordstretchfactor\fontdimen3\font minus
  \fontdimen4\font\relax}
\providecommand{\BIBforeignlanguage}[2]{{%
\expandafter\ifx\csname l@#1\endcsname\relax
\typeout{** WARNING: IEEEtran.bst: No hyphenation pattern has been}%
\typeout{** loaded for the language `#1'. Using the pattern for}%
\typeout{** the default language instead.}%
\else
\language=\csname l@#1\endcsname
\fi
#2}}
\providecommand{\BIBdecl}{\relax}
\BIBdecl

\bibitem{comon10ha}
P.~Comon and C.~Jutten, Eds., \emph{Handbook of Blind Source Separation -
  Independent Component Analysis and Applications}.\hskip 1em plus 0.5em minus
  0.4em\relax Academic Press, 2010.

\bibitem{avargel07mu}
Y.~Avargel and I.~Cohen, ``On multiplicative transfer function approximation in
  the short-time {F}ourier transform domain,'' \emph{IEEE Signal Processing
  Letters}, vol.~14, no.~5, pp. 337--340, 2007.

\bibitem{vincent10pr}
E.~Vincent, M.~G. Jafari, S.~A. Abdallah, M.~D. Plumbley, and M.~E. Davies,
  ``Probabilistic modeling paradigms for audio source separation,''
  \emph{Machine Audition: Principles, Algorithms and Systems}, pp. 162--185,
  2010.

\bibitem{hyvarinen01in}
A.~Hyv{\"a}rinen, J.~Karhunen, and E.~Oja, Eds., \emph{Independent Component
  Analysis}.\hskip 1em plus 0.5em minus 0.4em\relax Wiley and Sons, 2001.

\bibitem{Win07}
S.~Winter, W.~Kellermann, H.~Sawada, and S.~Makino, ``{MAP}-based
  underdetermined blind source separation of convolutive mixtures by
  hierarchical clustering and l1-norm minimization,'' \emph{EURASIP Journal on
  Advances in Signal Processing}, p. Article ID 24717, 2007.

\bibitem{mandel10mo}
M.~Mandel, R.~J. Weiss, D.~P. Ellis \emph{et~al.}, ``Model-based
  expectation-maximization source separation and localization,'' \emph{IEEE
  Trans. Audio, Speech, Lang. Process.}, vol.~18, no.~2, pp. 382--394, 2010.

\bibitem{liutkus11ga}
A.~Liutkus, B.~Badeau, and G.~Richard, ``Gaussian processes for underdetermined
  source separation,'' \emph{IEEE Trans. Signal Process.}, vol.~59, no.~7, pp.
  3155--3167, 2011.

\bibitem{ephraim84sp}
D.~Ephraim, Yariv~Malah, ``Speech enhancement using a minimum-mean square error
  short-time spectral amplitude estimator,'' \emph{IEEE Trans. Acoust., Speech,
  Signal Process.}, vol.~33, no.~6, pp. 443--445, 1984.

\bibitem{benaroya03no}
L.~Benaroya, L.~Donagh, F.~Bimbot, and R.~Gribonval, ``Non negative sparse
  representation for {W}iener based source separation with a single sensor,''
  in \emph{Proc. IEEE Int. Conf. Acoust., Speech, Signal Process. (ICASSP)},
  vol.~6, 2003, pp. 613--616.

\bibitem{benaroya06au}
L.~Benaroya, F.~Bimbot, and R.~Gribonval, ``Audio source separation with a
  single sensor,'' \emph{IEEE Trans. Audio, Speech, Lang. Process.}, vol.~14,
  no.~1, pp. 191--199, 2006.

\bibitem{fevotte05ma}
C.~F{\'e}votte and J.-F. Cardoso, ``Maximum likelihood approach for blind audio
  source separation using time-frequency {G}aussian source models,'' in
  \emph{Proc. IEEE Workshop Applicat. Signal Process. to Audio and Acoust.
  ({WASPAA})}, New Paltz, NJ, 2005.

\bibitem{ozerov10mu}
A.~Ozerov and C.~F{\'e}votte, ``Multichannel nonnegative matrix factorization
  in convolutive mixtures for audio source separation,'' \emph{IEEE Trans.
  Audio, Speech, Lang. Process.}, vol.~18, no.~3, pp. 550--563, 2010.

\bibitem{duong10un}
N.~Duong, E.~Vincent, and R.~Gribonval, ``Under-determined reverberant audio
  source separation using a full-rank spatial covariance model,'' \emph{IEEE
  Trans. Audio, Speech, Lang. Process.}, vol.~18, no.~7, pp. 1830--1840, 2010.

\bibitem{ozerov12ag}
A.~Ozerov, E.~Vincent, and F.~Bimbot, ``A general flexible framework for the
  handling of prior information in audio source separation,'' \emph{IEEE Trans.
  Audio, Speech Lang. Process.}, vol.~20, no.~4, pp. 1118--1133, 2012.

\bibitem{lee99le}
D.~Lee and H.~Seung, ``Learning the parts of objects by non-negative matrix
  factorization,'' \emph{Nature}, vol. 401, pp. 788--791, 1999.

\bibitem{lee01al}
------, ``Algorithms for non-negative matrix factorization,'' \emph{Advances in
  Neural Information Processing Systems}, vol.~13, pp. 556 -- 562, 2001.

\bibitem{fevotte09no}
C.~F{\'e}votte, N.~Bertin, and J.-L. Durrieu, ``Nonnegative matrix
  factorization with the {I}takura-{S}aito divergence. {W}ith application to
  music analysis,'' \emph{Neural Computation}, vol.~21, no.~3, pp. 793--830,
  2009.

\bibitem{yoshioka11bl}
T.~Yoshioka, T.~Nakatani, M.~Miyoshi, and H.~G. Okuno, ``Blind separation and
  dereverberation of speech mixtures by joint optimization,'' \emph{IEEE Trans.
  Audio, Speech, Lang. Process.}, vol.~19, no.~1, pp. 69--84, 2011.

\bibitem{anemuller99on}
J.~Anem{\"u}ller and T.~Gramss, ``On-line blind separation of moving sound
  sources,'' in \emph{Proc. Int. Conf. Independent Component Analysis and Blind
  Source Separation (ICA)}, Aussois, France, 1999.

\bibitem{koutras00bl}
A.~Koutras, E.~Dermatas, and G.~Kokkinakis, ``Blind speech separation of moving
  speakers in real reverberant environments,'' in \emph{Proc. IEEE Int. Conf.
  Acoust., Speech, Signal Process. (ICASSP)}, Istanbul, Turkey, 2000.

\bibitem{hild02bl}
K.~E. Hild~II, D.~Erdogmus, and J.~C. Principe, ``Blind source separation of
  time-varying, instantaneous mixtures using an on-line algorithm,'' in
  \emph{Proc. IEEE Int. Conf. Acoust., Speech, Signal Process. (ICASSP)},
  Orlando, Florida, 2002.

\bibitem{aichner03on}
R.~Aichner, H.~Buchner, S.~Araki, and S.~Makino, ``On-line time-domain blind
  source separation of nonstationary convolved signals,'' in \emph{Proc. Int.
  Conf. Independent Component Analysis and Blind Source Separation (ICA)},
  Nara, Japan, 2003.

\bibitem{prieto05bl}
R.~E. Prieto and J.~Pamornpol, ``Blind source separation for time-variant
  mixing systems using piecewise linear approximations,'' in \emph{Proc. IEEE
  Int. Conf. Acoust., Speech, Signal Process. (ICASSP)}, Philadelphia, PN,
  2005.

\bibitem{mukai03ro}
R.~Mukai, H.~Sawada, S.~Araki, and S.~Makino, ``Robust real-time blind source
  separation for moving speakers in a room,'' in \emph{Proc. IEEE Int. Conf.
  Acoust., Speech, Signal Process. (ICASSP)}, 2003.

\bibitem{addison06bl}
W.~Addison and S.~Roberts, ``Blind source separation with non-stationary mixing
  using wavelets,'' in \emph{Proc. Int. Conf. Independent Component Analysis
  and Blind Source Separation (ICA)}, Charleston, SC, 2006.

\bibitem{nakadai09so}
K.~Nakadai, H.~Nakajima, Y.~Hasegawa, and H.~Tsujino, ``Sound source separation
  of moving speakers for robot audition,'' in \emph{Proc. IEEE Int. Conf.
  Acoust., Speech, Signal Process. (ICASSP)}, Taipei, Taiwan, 2009.

\bibitem{araki07un}
S.~Araki, H.~Sawada, R.~Mukai, and S.~Makino, ``Underdetermined blind sparse
  source separation for arbitrarily arranged multiple sensors,'' \emph{Signal
  Process.}, vol.~87, no.~8, pp. 1833--1847, 2007.

\bibitem{loesch09on}
B.~Loesch and B.~Yang, ``Online blind source separation based on time-frequency
  sparseness,'' in \emph{Proc. IEEE Int. Conf. Acoust., Speech, Signal Process.
  (ICASSP)}, Taipei, Taiwan, 2009.

\bibitem{simon12ag}
L.~Simon and E.~Vincent, ``A general framework for online audio source
  separation,'' in \emph{Proc. Int. Conf. on Latent Variable Analysis and
  Signal Separation (LVA/ICA)}, Tel-Aviv, Israel, 2012.

\bibitem{markovich10su}
S.~Markovich-Golan, S.~Gannot, and I.~Cohen, ``Subspace tracking of multiple
  sources and its application to speakers extraction,'' in \emph{Proc. IEEE
  Int. Conf. Acoust., Speech, Signal Process. (ICASSP)}, Dallas, TX, 2010.

\bibitem{weinstein94it}
E.~Weinstein, A.~Oppenheim, M.~Feder, and J.~Buck, ``Iterative and sequential
  algorithms for multisensor signal enhancement,'' \emph{IEEE Trans. Signal
  Process.}, vol.~42, no.~4, pp. 846--859, 1994.

\bibitem{higushi12un}
T.~Higuchi, N.~Takamune, N.~Tomohiko, and H.~Kameoka, ``Underdetermined blind
  separation and tracking of moving sources based on {DOA-HMM},'' in
  \emph{Proc. IEEE Int. Conf. Acoust., Speech, Signal Process. (ICASSP)},
  Florence, Italy, 2014.

\bibitem{bishop06pa}
C.~Bishop, \emph{Pattern Recognition and Machine Learning}.\hskip 1em plus
  0.5em minus 0.4em\relax Springer, 2006.

\bibitem{gannot03on}
S.~Gannot and M.~Moonen, ``On the application of the unscented {K}alman filter
  to speech processing,'' in \emph{Proc. IEEE Int. Workshop on Acoustic Echo
  and Noise Control (IWAENC)}, Kyoto, Japan, 2003.

\bibitem{kounades15av}
D.~Kounades-Bastian, L.~Girin, X.~Alameda-Pineda, S.~Gannot, and R.~Horaud, ``A
  variational {EM} algorithm for the separation of moving sound sources,'' in
  \emph{Proc. IEEE Workshop Applicat. Signal Process. to Audio and Acoust.
  ({WASPAA})}, New Paltz, NJ, 2015.

\bibitem{neeser93pr}
F.~Neeser and J.~Massey, ``Proper complex random processes with applications to
  information theory,'' \emph{IEEE Trans. Info. Theory}, vol.~39, no.~4, pp.
  1293--1302, 1993.

\bibitem{virtanen07mo}
T.~Virtanen, ``Monaural sound source separation by nonnegative matrix
  factorization with temporal continuity and sparseness criteria,'' \emph{IEEE
  Trans. Audio, Speech, Lang. Process.}, vol.~15, no.~3, pp. 1066--1074, 2007.

\bibitem{mohammadiha13su}
N.~Mohammadiha, P.~Smaragdis, and A.~Leijon, ``Supervised and unsupervised
  speech enhancement using nonnegative matrix factorization,'' \emph{IEEE
  Trans. Audio, Speech, Lang. Process.}, vol.~21, no.~10, pp. 2140--2151, 2013.

\bibitem{parra00co}
L.~Parra and C.~Spence, ``Convolutive blind separation of non-stationary
  sources,'' \emph{IEEE Trans. Speech, Audio Process.}, vol.~8, no.~3, pp.
  320--327, 2000.

\bibitem{gannot01si}
S.~Gannot, D.~Burshtein, and E.~Weinstein, ``Signal enhancement using
  beamforming and nonstationarity with applications to speech,'' \emph{IEEE
  Trans. Signal Process.}, vol.~49, no.~8, pp. 1614--1626, 2001.

\bibitem{McLachlanEM}
G.~McLachlan and K.~Thriyambakam, \emph{The EM algorithm and extensions}.\hskip
  1em plus 0.5em minus 0.4em\relax New-York, USA: John Wiley and sons, 1997.

\bibitem{smidlVAR}
V.~Smidl and A.~Quinn, \emph{The Variational Bayes Method in Signal
  Processing}.\hskip 1em plus 0.5em minus 0.4em\relax Berlin: Springer-Verlag,
  2006.

\bibitem{hjorungnes07co}
A.~Hjorungnes and D.~Gesbert, ``Complex-valued matrix differentiation:
  Techniques and key results,'' \emph{IEEE Trans. Signal Process.}, vol.~55,
  no.~6, pp. 2740--2746, June 2007.

\bibitem{sturmel12li}
N.~Sturmel, A.~Liutkus, J.~Pinel, L.~Girin, S.~Marchand, G.~Richard, R.~Badeau,
  and L.~Daudet, ``Linear mixing models for active listening of music
  productions in realistic studio conditions,'' in \emph{Proc. Convention of
  the Audio Engineering Society (AES)}, Budapest, Hungary, 2012.

\bibitem{garofolo93ti}
J.~S. Garofolo, L.~F. Lamel, W.~M. Fisher, J.~G. Fiscus, D.~S. Pallett, N.~L.
  Dahlgren, and V.~Zue, ``Timit acoustic-phonetic continuous speech corpus,''
  1993, linguistic Data Consortium, Philadelphia.

\bibitem{hummersone13a}
C.~Hummersone, R.~Mason, and T.~Brookes, ``A comparison of computational
  precedence models for source separation in reverberant environments,''
  \emph{J. Audio Eng. Soc}, vol.~61, no. 7/8, pp. 508--520, 2013.

\bibitem{vincent06pe}
E.~Vincent, R.~Gribonval, and C.~F{\'e}votte, ``Performance measurement in
  blind audio source separation,'' \emph{IEEE Trans. Audio, Speech, Lang.
  Process.}, vol.~14, no.~4, pp. 1462--1469, 2006.

\bibitem{vincent07fi}
E.~Vincent, H.~Sawada, P.~Bofill, S.~Makino, and J.~Rosca, ``First stereo audio
  source separation evaluation campaign: data, algorithms and results,'' in
  \emph{Proc. Int. Conf. on Independent Component Analysis and Signal
  Separation (ICA)}, London, UK, 2007, pp. 552--559.

\bibitem{dorfan15tr}
Y.~Dorfan and S.~Gannot, ``Tree-based recursive expectation-maximization
  algorithm for localization of acoustic sources,'' \emph{IEEE/ACM Trans.
  Audio, Speech, Lang. Process.}, vol.~23, no.~10, pp. 1692--1703, 2015.

\bibitem{may11ap}
T.~May, S.~Van De~Par, and A.~Kohlrausch, ``A probabilistic model for robust
  localization based on a binaural auditory front-end,'' \emph{IEEE Trans.
  Audio, Speech, Lang. Process.}, vol.~19, no.~1, pp. 1--13, 2011.

\bibitem{woodruff12bi}
J.~Woodruff and D.~Wang, ``Binaural localization of multiple sources in
  reverberant and noisy environments,'' \emph{IEEE Trans. Audio, Speech, Lang.
  Process.}, vol.~20, no.~5, pp. 1503--1512, 2012.

\bibitem{traa14mu}
J.~Traa and P.~Smaragdis, ``Multichannel source separation and tracking with
  {RANSAC} and directional statistics,'' \emph{IEEE/ACM Trans. Audio, Speech,
  Lang. Process.}, vol.~22, no.~12, pp. 2233--2243, 2014.

\bibitem{dorfan15sp}
Y.~Dorfan, D.~Cherkassky, and S.~Gannot, ``Speaker localization and separation
  using incremental distributed expectation-maximization,'' in \emph{Proc.
  Europ. Signal Process. Conf. (EUSIPCO)}, Nice, France, 2015, pp. 1256--1260.

\bibitem{allen79im}
J.~B. Allen and D.~A. Berkley, ``Image method for efficiently simulating
  small-room acoustics,'' \emph{The Journal of the Acoustical Society of
  America}, vol.~65, no.~4, pp. 943--950, 1979.

\end{thebibliography}
%

\end{document}